\documentclass[useAMS,usenatbib]{mn2e}

\newcommand{\kms}{ km s$^{-1}$\xspace}
\newcommand{\ang}{\mbox{\AA}\xspace}
\newcommand{\nhI}{N$_{\rm H \ I}$\xspace}
\newcommand{\w}{W$_{0}$}
\newcommand{\za}{z$_{abs}$}
\newcommand{\lala}{$\lambda\lambda$\xspace}
\newcommand\Lya{Lyman-$\alpha$\xspace}

\usepackage{graphics}
\usepackage{graphicx}
\usepackage{colordvi}
\usepackage{epsfig}
\usepackage{lscape}
\usepackage{rotating}
\usepackage{xspace}
\usepackage{float}
\begin{document}

\title[A Sample of Sub-DLAs at z $\la$ 1.5.]{A MIKE + UVES Survey of Sub-Damped Lyman-$\alpha$ Systems at $z < 1.5$. }
\author[J. D. Meiring, V.P. Kulkarni et al.]{Joseph D. Meiring$^{1}$\thanks{Previous Address: Department of Physics and Astronomy, University of South Carolina, Columbia SC, 29208}, 
James T. Lauroesch$^{1}$,
Varsha P. Kulkarni$^{2}$, 
Celine P\'eroux$^{3}$,
\newauthor  Pushpa Khare$^{4}$, $\&$ Donald G. York$^{5,6}$ \\
$^{1}$Department of Physics and Astronomy, University of Louisville, Louisville, Ky 40292 USA\\
$^{2}$Department of Physics and Astronomy, University of South Carolina, Columbia, SC 29208, USA \\
$^{3}$Laboratoire d'Astrophysique de Marseille, OAMP, Universite Aix-Marseill\`e $\&$ CNRS, 13388 Marseille cedex 13, France \\
$^{4}$Department of Physics, Utkal University, Bhubaneswar, 751004, India \\
$^{5}$Department of Astronomy and Astrophysics, University of Chicago, Chicago, IL 60637, USA \\ 
$^{6}$Enrico Fermi Institute, University of Chicago, Chicago, IL 60637, USA \\ }

\date{Accepted ... Received ...; in original form ...}

\pagerange{\pageref{firstpage}--\pageref{lastpage}} \pubyear{}

\maketitle

\label{firstpage}

\begin{abstract}

We have combined the results from our recent observations of Damped and sub-Damped \Lya systems with the MIKE and UVES spectrographs
on the Magellan Clay and VLT Kueyen  telescopes with ones from the literature to determine the 
\nhI-weighted mean metallicity of these systems based both on Fe, 
a depleted element in QSO absorbers and the local ISM,
and Zn a relatively undepleted element. In each case, the \nhI weighted mean metallicity is higher and shows faster 
evolution in sub-DLAs than the classical DLA systems.
Large grids of photoionisation models over the sub-DLA \nhI range with \textsc{cloudy} show that the 
ionisation corrections to the abundances are in general small, however the fraction of ionized H can be up to
$\sim$90 per cent. The individual spectra have been shifted to the rest frame of the absorber and averaged together to determine
the average properties of these systems at $z<1.5$. We find that the average abundance pattern of the Sub-DLA systems is similar to the gas in the halo of the
Milky Way, with an offset of $\sim$0.3 dex in the overall metallicity. Both DLAs and Sub-DLAs show similar characteristics 
in their relative abundances patterns, although the DLAs have smaller $\langle$[Mn/Zn]$\rangle$ as well as higher $\langle$[Ti/Zn]$\rangle$ and $\langle$[Cr/Zn]$\rangle$. 
We calculate the contribution of sub-DLAs to the metal budget of the Universe, and
find that the sub-DLA systems at $z<1.5$ contain a comoving density of metals $\Omega_{met}\sim$(3.5-15.8)$\times10^{5}$ M$_{\sun}$ Mpc$^{-3}$, at least twice
the comoving density of metals in the DLA systems. 
The sub-DLAs do however track global chemical evolution models much more closely than do the DLAs, perhaps indicating that 
they are a less dust biased metallicity indicator of galaxies at high redshifts than the DLA systems. 

\end{abstract}


\section{Introduction}
Even with the recent introduction of large telescopes coupled with sensitive instrumentation, studying the  properties of distant, high redshift ($z$) galaxies still remains a
difficult task. When studied via emission, biases towards more luminous galaxies, i.e. the Malmquist bias, are often present. 
One of the most fundamental and basic predictions of galaxy evolution models is that the metallicity of the galaxy should increase over time with the 
ongoing cycles of star formation and death that pollute the Interstellar Medium (ISM) of the galaxy with heavy elements in successively higher amounts with each subsequent generation of stars. 

Chemical evolution studies of the Milky Way are able to utilize individual stellar abundances over a wide range of metallicities to constrain models of the evolution 
of the Galaxy (see for instance \citet{PT98, Sam98, Nis00, Niss04}). Direct stellar abundance measurements are difficult in even the
most nearby galaxies, and currently impossible at higher redshifts. 
An opportunity to study the ISM of high redshift galaxies occurs if a galaxy is fortuitously
aligned along the line of sight to a distant Quasar (QSO), as the gas produces narrow absorption lines
in the continuum of the background QSO. 

These absorption line systems seen in the spectra of QSOs provide an alternative approach to studying distant galaxies,
and offer a unique window  
into the ISM of high $z$ galaxies  as well as the diffuse Intergalactic Medium (IGM) (see for instance \citet{Leh07, Tripp08, Dan08, FMY08}
for recent studies of the metal content of the IGM). As these galaxies are not selected upon any morphological criteria, this approach 
studies galaxies independent of a particular morphological type. 
The metal (Z$>$2) abundances in the ISM of these galaxies yield important clues into the processes 
of star formation and death and the overall chemical enrichment of the system being investigated.

Based upon their neutral hydrogen column densities, intervening absorption line systems with strong \Lya 
lines showing damping wings can be classified into the Damped \Lya systems (DLAs with log \nhI$\ge$20.3) and 
sub-DLA systems (sub-DLA 19 $\la$ log \nhI $<$ 20.3, \citealt{Per01}).
The damping wings which give rise to the name of these systems and
can be used to accurately measure \nhI in these systems begin the sub-DLA regime of log \nhI$\sim$19.0. 
With their high gas content, the DLA and sub-DLA systems are expected to be associated directly with galaxies at high redshift, and 
several systems have been confirmed with followup imaging at lower $z$ \citep{LeBrun97, BTJ01, Chen03}.

After much investigation, the DLA systems have
been found to have low metallicity ($\sim$0.15Z$_{\sun}$) and show little chemical evolution from $0.1<z<3.5$ \citep{Kul05, Kul07, Mei06, Pro03, WGP05}.
which contradicts galactic chemical evolution models \citep{PFH99, Cen03}. If the DLAs, which do not appear to contain the metals predicted by 
chemical evolution models, could these missing metals be hiding in other locations, have they been systematically underestimated in DLAs, or both?
This survey was conducted to obtain high resolution spectra in a homogenous sample of the lesser studied sub-DLAs to further investigate their role in cosmic chemical evolution. 

The global metal budget of the Universe can be estimated based on the integrated comoving star formation rate density $\dot{\rho}_\star$, and the 
theoretical metal production yields of stars $\langle$p$_{Z}\rangle$ \citep{Searle72, Bou08}. 
Adding the contributions from the \Lya forest, Lyman break galaxies, sub-mm galaxies,  BX/BM galaxies (i.e. \citealt{Adel04}) and the 
DLA systems to the global metal budget of the 
Universe, a dearth of $\sim40\%$ of the expected metal content is seen \citep{BLP06, Bou08}. 
Several groups have reported high abundances in sub-DLA systems \citep{Kh04, Per06a, Pro06, Mei07, Sri08, QRB08}. 
Strong Lyman-$\alpha$ forest lines with log \nhI$\la$16 have also been recently observed to have super-solar metallicities,
although ionisation correction factors are important and not easily constrained \citep{Char03, Ding03, Mas05}. 
It has also been suggested that a large portion of these missing metals may 
be locked into the hot intergalactic medium as well \citep{Ferr05}.

Although redshifts $z < 1.5$ span ~70$\%$ of the age of the Universe (assuming the ``737'' concordance 
cosmology of $\Omega_{\Lambda}=0.7$,  $\Omega_{m}=0.3$, H$_{0}$=70 \kms Mpc$^{-1}$ which we
utilize throughout this work), few observations have been made of $z < 1.5$ sub-DLAs  
due to the lack of spectrographs with enough sensitivity in short wavelengths and the 
the paucity of known sub-DLAs in this redshift range since \nhI must be determined with space based spectrographs as the \Lya line lies in the UV at these redshifts. 
This range is clearly important for understanding the nature of sub-DLA systems along with galactic chemical evolution.

In this paper we discuss the full sample of sub-DLAs obtained in this survey with the Magellan-II telescope and the MIKE spectrograph and the VLT Kuyuen telescope with 
the UVES spectrograph. The structure of this paper is as follows. 
In $\S$ \ref{Sec:Obs} we discuss details of our sample. In $\S$ 3, we discuss the abundances and column density determinations. $\S$ \ref{Sec:Cloudy} gives information 
on photoionisation models with \textsc{cloudy}. In $\S$ 5 we discuss the kinematical properties of these systems. 
In $\S$ \ref{Sec:Weighted} we give the 
\nhI-weighted mean metallicity based on both Fe and Zn. We discuss the averaged sub-DLA spectrum in $\S$ \ref{Sec:Averaged}. In $\S$ \ref{Sec:Global} we estimate the contribution of sub-DLA systems to the
comoving metal budget, and in $\S$ \ref{Sec:Discussion} we discuss the implications of these measurements. 

%%%%%%%%%%%%%%%%%%%%%%%%%%%%%%%%%%%%%%%%%%%%%%%%%%%%%%%%%%%%%%%%%%%%%%%%%%%%%%%%%%%%%%%%%%%%%%%%%%%%%%%%%%%%%%%%
%%%%%%%%%%%%%%%%%%%%%%%%%%%%%%%%%%%%%%%%%%%%%%%%%%%%%%%%%%%%%%%%%%%%%%%%%%%%%%%%%%%%%%%%%%%%%%%%%%%%%%%%%%%%%%%%

\section{Sample Selection} \label{Sec:Obs}
The observations in this sample were made with the 6.5m Magellan-II Clay telescope and the Magellan Inamori Kyocera Echelle (MIKE) spectrograph (see \citealt{Bern03}) 
along with the 8.2m Very Large Telescope (VLT) Kueyen telescope with the Ultraviolet Visual Echelle Spectrograph
(UVES, see \citealt{Dodor00} for details of the instrument). These observations were undertaken from 2005 until 2008. Our total
sample includes spectra of 27 QSOs and a total of 34 systems, with 31 sub-DLAs or strong Lyman-limit systems and three DLAs. 
These results have been published in \citet{Mei07, Mei08, Mei09} and \citet{Per08}.

Our targets were  selected from the \citet{Rao06} catalog of \textit{Mg II absorbers} identified in Hubble Space Telescope (HST) spectra having
0.65 $\la$ \za $\la$ 1.5 (so that the Zn II \lala 2026,2062 lines were redhsifted into the optical) with 19.0$\la$\nhI$<$20.3. 
Due to the lower H I column densities and hence lower expected Zn II column densities for sub-solar metallicity systems in the \nhI range, 
we required the background QSO to be brighter than $m_{g}\la18.5$ in order to achieve S/N ratios adequate to detect the expected weak Zn II lines.
 
Ideally, a blind survey of QSOs is the best possible option with minimal bias. 
This is however unfeasible due to the necessity of both ground and space based spectra, and
the small coincidence rate of Sub-DLAs or DLAs in the spectra of background QSOs. 
An Mg II selection criterion is not likely 
to significantly bias the sample, as the rest frame equivalent width of the Mg II \lala 2796, 2803 lines 
does not correlate well with [Zn/H] or [Fe/H] for DLAs or sub-DLAs. \citep{Kul07, Des09}. 
Of the $\sim$45 systems from the \citet{Rao06} sample, only eight systems remain at z$\la$1.5
that fit the selection criteria above and not observed in the sample presented here due to scheduling and time constraints. 
These objects have similar mean W$_{0}^{2796}$ ($\sim$1.4 \ang compared to $\sim$1.6 \ang for systems that we have observed), and thus
are unlikely to bias the sample. 

\begin{figure} 
\begin{center}
\includegraphics[angle=90, width=\linewidth]{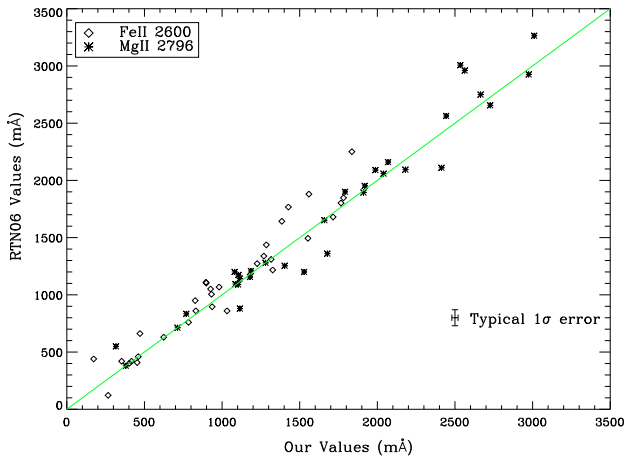}
\caption[Comparison of Equivalent Widths]{ A comparison of the rest frame equivalent widths of Mg II $\lambda$ 2796 and Fe II $\lambda$ 2600 
from these data, and from the lower resolution spectra from \citet{Rao06}. \label{Fig:EWCompare}}
\end{center}
\end{figure}

Throughout this paper the QSO names are given in J2000 coordinates.
We show a comparison of the rest frame equivalent widths as measured by \citet{Rao06} 
and the values determined in these higher resolution spectra for the Mg II $\lambda$ 2796 and Fe II $\lambda$ 2600 
lines in Figure \ref{Fig:EWCompare}. The values appear to be in excellent agreement, with no systematic
differences in the two measurements.

\section{Abundances } \label{Sec:Abund}

Total abundances for the observed systems are given in Table \ref{Tab:Abund1}. These values have been given in our previous works (see \citealt{Mei07, Mei08, Mei09, Per06b}),
and are compiled here for completeness and convenience to the reader. 
We have used the total column densities (i.e. the sum of the column densities in the individual components of a system that were determined via profile fitting method)
along with the total \nhI as given in Table \ref{Tab:Abund1}, to determine the abundances of these systems. 
We have not placed any ionisation corrections while determining these abundances (the issue of ionisation corrections is discussed further in Section \ref{Sec:Cloudy}),
and have assumed the first ions to be the dominant 
ionisation species of the elements for which these abundances have been determined, namely Zn, Fe, Mn, Cr, and Si. 
Solar systems abundances from \citet{Lodd03} are also given in Table \ref{Tab:Abund1}. 

In Table \ref{Tab:Abund1}, we have given the  metallicities
of Zn and Fe, the relative abundances of a several species, as well as the column
density ratios of adjacent ions of a few elements. All these are based on the
column densities obtained from the profile fitting analysis. Relative
abundances like [Zn/Fe], are indicators of dust depletion, and ion ratios like
Al II/Al III, may provide information about the ionization in the systems.

\setlength{\tabcolsep}{3pt}
\begin{sidewaystable*}
\setcounter{table}{1}
\scriptsize
\begin{center}
%\caption{Abundances for the absorbers in this sample. \label{Tab:Abund1}}
\begin{tabular}{lcccccccccccccc}
\hline
\hline	
QSO		&	$z_{abs}$&	log \nhI	&	[Zn/H]		&	[Fe/H]			&	[Fe/Zn]		&	[Si/Fe]			&	[Ca/Fe]		&		[Cr/Fe]	&	[Mn/Fe]		&	Al III / Al II	&Mg II / Mg I		&	Mg II / Al III		&	Fe II / Al III	&	$\Delta v_{90}$		\\
\hline
[X/Y]$_{\sun}$	&		 &			&	$-$7.37		&	$-$4.53			&	+2.84		&	+0.07			&	$-$1.13		&	$-$1.82		&	$-$1.97		&			&			&				&			&	km s$^{-1}$		\\
\hline																							
Q0005+0524	&	0.8514	&	19.08$\pm$0.04	&	$<-$0.47	&	$-$0.76$\pm$0.05	&	$>-$0.29	&		-	&	-			&	$<-$0.09	&	$<-$0.81	&	-		&	$>$2.14		&	$>$+1.21		&	0.68$\pm$0.04	&	39		\\
Q0012$-$0122	&	1.3862	&	20.26$\pm$0.02	&	$<-$1.34	&	$-$1.49$\pm$0.03	&	$>-$0.15	&	+0.12$\pm$0.08	&	-			&	$<-$0.53	&	$<-$0.86	&	$<-$0.19	&	$>$2.36		&	$>$+1.20		&	1.35$\pm$0.03	&	32		\\
Q0021+0104	&	1.3259	&	20.04$\pm$0.11	&	$<-$1.19	&	$-$0.82$\pm$0.11	&	$>+$0.37	&	$>$+0.14	&	-			&	$<-$0.66	&	$<-$0.85	&	-		&	$>$2.70		&	-			&	-		&	185		\\
%Q0021+0104B	&	1.5756	&	20.48$\pm$0.15	&	$<-$1.16	&	$-$1.34$\pm$0.15	&	$>-$0.18	&	+0.20$\pm$0.04	&	-			&	$<-$0.21	&	$<-$0.74	&	-		&	$>$1.96		&	$>$+2.31		&	2.35$\pm$0.08	&	227		\\
Q0123$-$0058  	&	1.4094	&	20.08$\pm$0.09	&	$-$0.45$\pm$0.20&	$-$0.55$\pm$0.12	&	$-$0.10$\pm$0.11&	+0.42$\pm$0.06	&	-			&	$-0.25\pm0.10$	&	$-0.40\pm0.12$	&	-		&	-		&	-			&	1.22$\pm$0.03	&	140		\\
Q0132$+$0116  	&	1.2712	&	19.70$\pm$0.09	&	$<-$0.54	&	$-$0.53$\pm$0.12	&	$>+$0.01	&	-		&	-			&	$<-0.67$	&	$<-1.10$	&	-		&	-		&	-			&	1.04$\pm$0.02	&	207		\\
Q0138$-$0005  	&	0.7821	&	19.81$\pm$0.09	&	+0.28$\pm$0.16  &	$>-$0.09		&	$>-$0.37	&	-		&	-			&	-		&	-		&	-		&	$>$2.91		&	-			&	-		&	95		\\	
Q0153$+$0009  	&	0.7714	&	19.70$\pm$0.09	&	$<-$0.34	&	 - 			&	-		&	-		&	-			&	-		&	-		&	-		&	$>$2.88		&	$>$+3.16		&	-		&	278		\\	
Q0354$-$2724	&	1.4051	 &  20.18$\pm$0.15	&	$-$0.08$\pm$0.16&	$-$0.50$\pm$0.16	& $-$0.43$\pm$0.06	&	-		&	-			&	+0.02$\pm$0.07	&	$-0.28\pm$0.08	&	$<$-0.79	&	$>$2.36		&	$>$1.67			&	1.74$\pm$0.06	&	176		\\
Q0427$-$1302	&	1.4080	&	19.04$\pm$0.04	&	$<-$0.58	&	$-$1.15$\pm$0.04	&	$>-$0.57	&	+0.16$\pm$0.03	&	-			&	$<$+0.33	&	$<$+0.26	&	$<-$1.14	&	-		&	-			&	$>$2.30		&	42		\\
%Q0449$-$1645  	&	1.0072	&	20.98$\pm$0.07	&	$-$0.96$\pm$0.08&	$-$1.34$\pm$0.08	&	$-$0.38$\pm$0.07&	+0.70$\pm$0.05	&	-			&	+0.20$\pm$0.02	&	$-0.21\pm0.04$	&	-		&	-		&	-			&	1.51$\pm$0.03	&	201		\\
Q0826$-$2230	&	0.9110	 &  19.04$\pm$0.04	&	+0.68$\pm$0.08	&	$-$0.94$\pm$0.06	& $-$1.63$\pm$0.08	&	$<$+0.58	&	$-$0.69$\pm$0.07	&	$<$+0.36	&	$<-$0.23	&	-		&	$>$1.67		&	$>$2.04			&	$>$1.85		&	276		\\
Q1009$-$0026A	&	0.8426   &  20.20$\pm$0.06	&	$<-$0.98	&	$-$1.28$\pm$0.07	&	$>-$0.31	&	-		&	-			&	$<-$0.16	&	$-0.16\pm$0.06	&	-		&	$>$2.57		&	$>$1.57			&	1.66$\pm$0.06	&	36		\\
Q1009$-$0026B	&	0.8866	 &  19.48$\pm$0.05	&	+0.25$\pm$0.06	&	$-$0.58$\pm$0.06	&   $-$0.83$\pm$0.06	&	$<-$0.18	&	$-$0.98$\pm$0.06	&	$<-$0.44	&	$-$0.05$\pm$0.08&	-		&	$>$2.11		&	$>$1.40			&	1.25$\pm$0.05	&	94		\\
Q1010$-$0047	&	1.3270	 &  19.81$\pm$0.05	&	$<-$0.75	&	$-$0.75$\pm$0.06	&	$>+$0.01	&	+0.42$\pm$0.07	&		-		&	$<-$0.34	&	$<-$0.71	&	$<-$0.71	&	$>$2.71		&	$>$1.69			&	1.25$\pm$0.05	&	156		\\
Q1037+0028	& 1.4244 	& 	20.04$\pm$0.12 	&	$<-0.63$	& 	$-$0.67$\pm$0.12 	&	$>-0.04$	& 	+0.18$\pm$0.04	&	 	-	 	&	$<-0.76$	&$-$0.34$\pm$0.05 	&	$<-0.96$	&	 -	 	&	$>$2.29		 	&	1.65$\pm$0.03	&	169		\\
Q1054$-$0020A	& 0.8301 	& 	18.95$\pm$0.18 	&	$<+0.18$	& 	$-$0.09$\pm$0.18 	&	$>-0.28$	&	 	-	&	 	-	 	&	$<-0.04$	&$-$0.09$\pm$0.07 	&	 	- 	&	$>$2.28	 	&	$>$1.24		 	&	 0.74$\pm$0.06 	&	85		\\
Q1054$-$0020B	& 0.9514 	& 	19.28$\pm$0.02 	&	$<-0.21$	& 	$-$1.09$\pm$0.02 	&	$>-0.88$	&	$<+0.52$	&	 	-	 	&	$<+0.36$	&	$<+0.23$	&	 	- 	&	$>$1.68	 	&	$>$1.61		 	&	$>$1.55		&	160		\\
Q1215$-$0034	& 1.5543 	& 	19.56$\pm$0.02 	&	$<-0.56$	& 	$-$0.64$\pm$0.02 	&	$>-0.08$	&	 	-	&	 	-	 	&	$<-0.20$	&	$<-0.79$	&	 	- 	&	-	 	&	$>$2.57		 	&	1.62$\pm$0.03	&	81		\\
Q1220$-$0040	& 0.9746 	& 	20.20$\pm$0.07	&	$<-1.14$	& 	$-$1.33$\pm$0.07 	&	$>-0.19$	&	 	-	&	 	-	 	&	$<-0.14$	&	$<-0.42$        &	 	- 	&	$>$2.69	 	&	$>$2.19		 	&	 1.72$\pm$0.06	&	95		\\
%Q1224+0037A	&	1.2346	 &  20.88$\pm$0.05	&	$<-$1.62	&	$>-$1.24		&	$>+$0.37	&			&		-		&	$<-$0.17	&	-		&	$<-$0.49	&	$>$2.52		&	$>$1.73			&	$>$1.98		&	94		\\
Q1224+0037	&	1.2665	 &  20.00$\pm$0.07	&	$<-$0.78	&	$-$0.93$\pm$0.11	&	$>-$0.16	&	$<-$0.31	&		-		&	$<-$0.35	&	$<-$0.41	&	$<-$0.74	&	$>$2.97		&	$>$1.99			&	1.56$\pm$0.12	&	280		\\
Q1228+1018	& 0.9376 	& 	19.41$\pm$0.02 	&	 $<-0.37$	& 	$-$0.30$\pm$0.02	&	$>+0.07$	&	 	-	&	 	-	 	&	$<-0.55$	&	$<-0.47$	&	 	- 	&	$>$2.29	 	&	$>$1.51		 	&	1.48$\pm$0.10	&	116		\\
Q1330$-$2056	& 0.8526 	& 	19.40$\pm$0.02 	&	$<-0.07$	& 	$-$1.07$\pm$0.02 	&	$>-1.00$	&	 	-	&	 	- 		&	$<+0.20$	&	$<-0.22$	&	 	- 	&	$>$1.88	 	&	$>$1.62		 	&	 1.21$\pm$0.05	&	578		\\
Q1436$-$0051A	& 0.7377 	& 	20.08$\pm$0.11 	&	$-$0.05$\pm$0.12& 	$-$0.61$\pm$0.11 	& $-$0.58$\pm$0.04	&	 	-	&	 $-$0.98$\pm$0.03 	&	$<-0.41$	& 	+0.04$\pm$0.03 	&	 	- 	&	-	 	&	-		 	&	- 		&	71		\\
Q1436$-$0051B	& 0.9281 	& 	$<$18.8		&	$>$+0.86	& 	$>-$0.07	 	& $-$0.94$\pm$0.06 	& 	+0.75$\pm$0.05 	&	 $-$0.79$\pm$0.04 	&	$<+0.20$	& 	+0.22$\pm$0.05 	&	 	- 	&	$>$2.63	 	&	$>$1.72		 	&	0.80$\pm$0.03   &	62		\\
Q1455$-$0045	& 1.0929 	& 	20.08$\pm$0.06	&	$<-0.80$	& 	$-$0.98$\pm$0.06 	&	$>-0.18$	& 	+0.00$\pm$0.10 	&	 	-	 	&    $-$0.35$\pm$0.13	& 	-0.51$\pm$0.15 	&	$<-$0.18 	&	$>$2.39	 	&	$>$1.27		 	&	0.92$\pm$0.02	&	121		\\
Q1631+1156	&	0.9004	&	19.70$\pm$0.04	&	$<-$0.15	&	$-$1.06$\pm$0.06	&	$>-$0.91	&	-		&	$-$0.85$\pm$0.07	&	$<$+0.36	&	$<$+0.40	&	-		&	$>$1.83		&	-			&	-		&	58		\\
Q2051+1950	&	1.1157	&	20.00$\pm$0.15	&	$+$0.27$\pm$0.18&	$-$0.45$\pm$0.15	&   $-$0.72$\pm$0.10	&	+0.06$\pm$0.07	&	$-$1.34$\pm$0.04	&	$-$0.31$\pm$0.10&	+0.16$\pm$0.03	&	$<-$0.17	&	$>$1.97		&	$>$+1.08		&	1.49$\pm$0.04	&	113		\\
Q2331+0038	&	1.1414	 &  20.00$\pm$0.05	&	$-$0.51$\pm$0.12&	$-$1.09$\pm$0.06	&	$-$0.59$\pm$0.11&	$<-$0.12	&		-		&	$<-$0.19	&	$<-$0.62	&	$<-$0.40	&	$>$2.31		&	$>$1.77			&	1.36$\pm$0.09	&	137		\\
Q2335$+$1501	&	0.6798	&	19.70$\pm$0.30	&	$+$0.07$\pm$0.34& 	$-$0.32$\pm$0.33	&	$-$0.39$\pm$0.05&	-		&	$-$1.29$\pm$0.04	&	$-0.12\pm0.11$	&	-		&	-		&	$>$2.77		&	-			&	-		&	95		\\	
Q2352$-$0028A	&	0.8730	&	19.18$\pm$0.09	&	$<-$0.14	&	$-$1.17$\pm$0.09	&	$>-$1.03	&	-		&	-			&	$<$+0.71	&	$<-$0.05	&	-		&	$>$2.42		&	-			&	-		&	120		\\
Q2352$-$0028B	&	1.0318	&	19.81$\pm$0.13	&	$<-$0.51	&	$-$0.37$\pm$0.13	&	$>+$0.14	&	+0.51$\pm$0.03	&	-			&	$-$0.13$\pm$0.06&	$<-$1.07	&	-		&	$>$2.41		&	$>$+1.53		&	1.50$\pm$0.03	&	164		\\
Q2352$-$0028C	&	1.2467	&	19.60$\pm$0.24	&	$<-$0.7		&	$-$0.86$\pm$0.24	&	$>-$0.16	&	-		&				&	$<-$0.22	&	$<-$0.64	&	$<-$0.10	&	$>$2.91		&	$>$+1.85		&	0.82$\pm$0.03	&	240		\\
\hline
\end{tabular}
\end{center}
\begin{minipage}{200mm}
\textbf{Table 1:} Elemental abundances and adjacent ion ratios for the absorbers in this sample. The solar abundances from \citet{Lodd03} are given in the headings as well. In the last four columns, the ratios
of the column densities of several pairs of ions are given in dex. \nhI is given in logarithmic form in units of cm$^{-2}$. \label{Tab:Abund1}
\end{minipage}
\end{sidewaystable*}

%%%%%%%%%%%%%%%%%%%%%%%%%%%%%%%%%%%%%%%%%%%%%%%%%%%%%%%%%%%%%%%%%%%%%%%%%%%%%%%%%%%%%%%%%%%%%%%%
%%%%%%%%%%%%%%%%%%%%%%%%%%%%%%%%%%%%%%%%%%%%%%%%%%%%%%%%%%%%%%%%%%%%%%%%%%%%%%%%%%%%%%%%%%%%%%%%

\section{Photoionisation Models } \label{Sec:Cloudy}
The gas in DLA systems is expected to remain largely neutral due to self-shielding of the UV photons capable of ionization. 
The ionisation correction factor, defined here as
\begin{equation}
\epsilon=[X/H]_{\rm total} - [X^{+}/H^{0}] 
\end{equation}
\noindent where the total column densities include contributions from all ionisation stages, has been investigated for DLAs by several groups 
\citep{Howk99, Vlad01, Pro02}. The ionisation corrections for most elements in DLA systems were found to be
$\la$0.2 dex in most cases. Due to the lesser amount of H I in the sub-DLA systems, and hence less shielding, it is 
expected that they should show a greater amount of ionisation. 

    It has previously been shown that the ionisation corrections based on \textsc{cloudy} models are in general also expected to be small for sub-DLA systems \citep{Des03, Mei07, Mei08}. 
Here, we expand upon our previous \textsc{cloudy} modeling with grids of photoionisation models over the entire sub-DLA \nhI range. 
Models were created with version C06.02.b of \textsc{cloudy}, last described by \citet{Fer98}. Grids of \textsc{cloudy} models were computed assuming 
that the spectrum of ionizing radiation striking the cloud followed the form of the extra-galactic UV background of Haardt $\&$ Madau \citep{HM96, MHR99} 
at the appropriate redshift of the absorber, plus the model stellar atmosphere of Kurucz with a temperature of 30,000K to simulate a 
radiation field produced via an O/B-type star. Both radiation fields were combined in equal parts for the final incident spectrum.
Plots of both types of spectra can be found in Hazy, the documentation for \textsc{cloudy} 
which can be found at \verb http://www.nublado.org . 
In addition, our \textsc{cloudy} models have included a cosmic ray background and cosmic microwave background. These simulations
however do not include radiation from local shocks caused by supernovae, or compact sources such as white dwarfs or compact binary systems, all of which 
likely contribute the ionizing radiation field.  
	
For each of the grids of models, the ionisation parameter defined by
\begin{equation}
U=\frac{n_{\gamma}}{n_{H}}=\frac{\Phi_{912}}{cn_{H}}
\end{equation}
\noindent (where $\Phi_{912}$ is the flux of radiation with h$\nu$ $>$ 13.6 eV) was increased from log U=$-$6.0 to 0. 

\begin{figure}
\begin{center}
\includegraphics[angle=90, width=\linewidth]{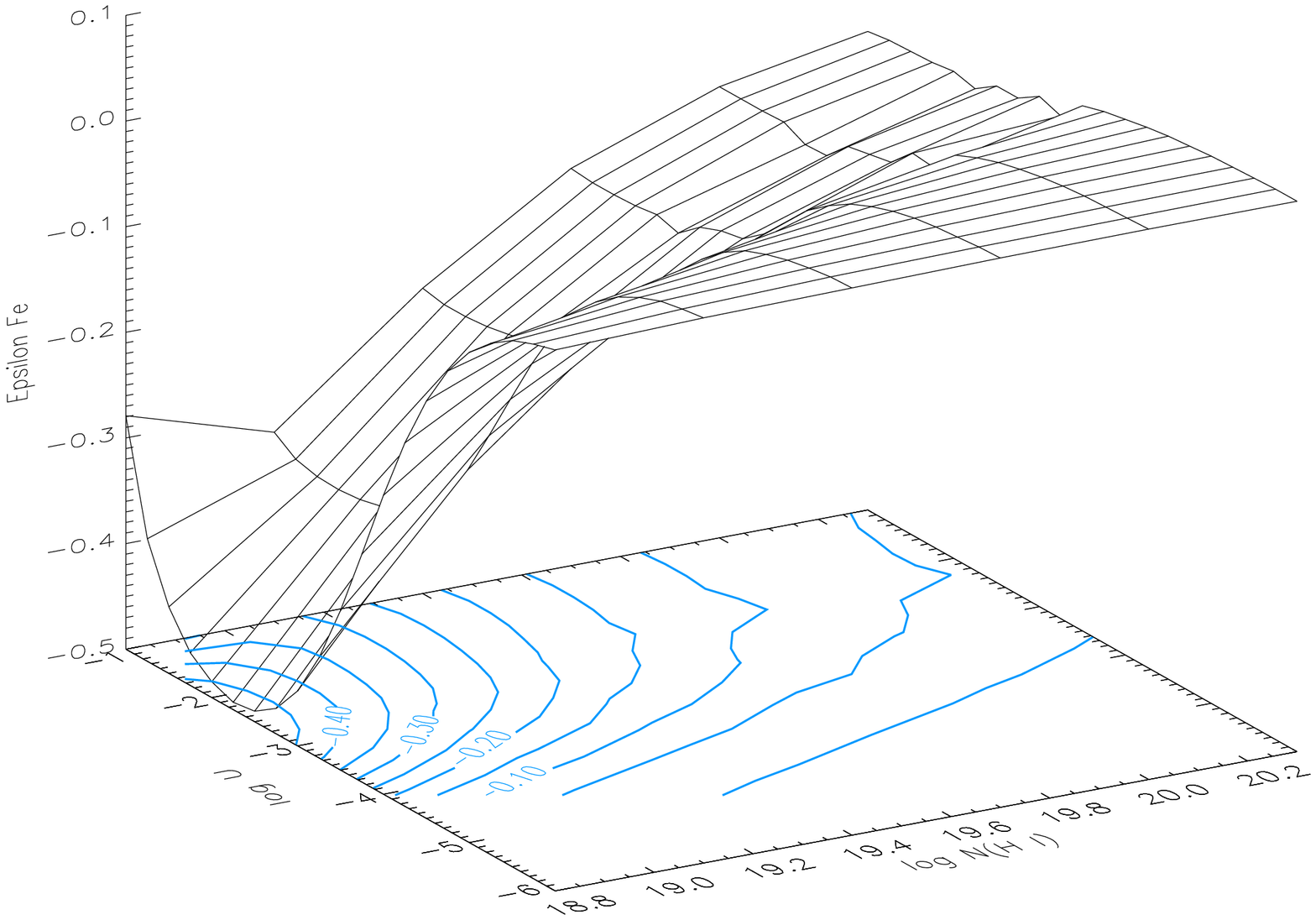}
\caption{ Results of the \textsc{cloudy} photoionisation models for the ionisation correction needed for Fe as a function of log U and \nhI. The blue
contours on the horizontal plane represent the ionisation corrections in steps of 0.05 dex. \label{Fig:Fe_cor}}
\includegraphics[angle=90, width=\linewidth]{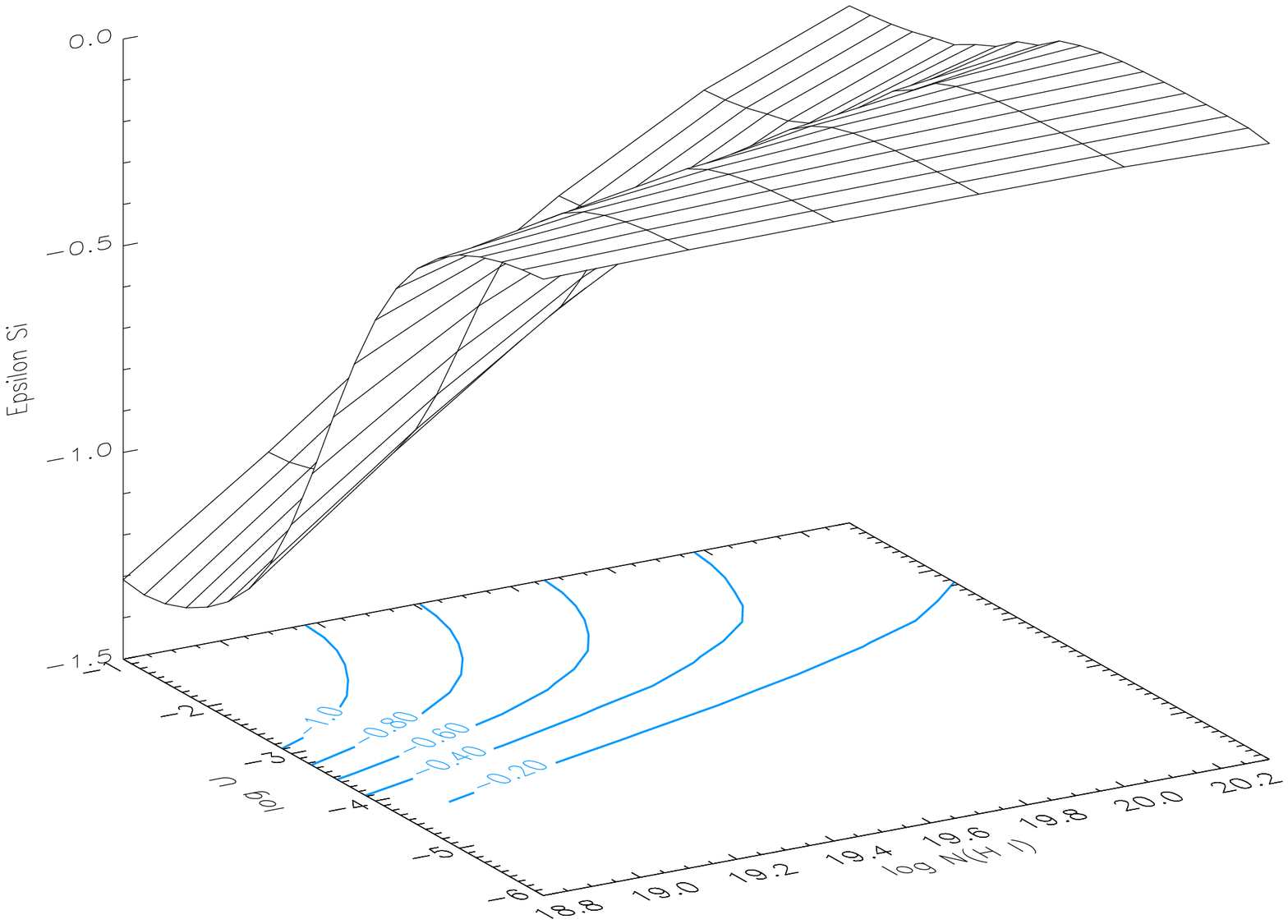}
\caption{ Results of the \textsc{cloudy} photoionisation models for the ionisation correction needed for Si as a function of log U and \nhI. The blue
contours on the horizontal plane represent the ionisation corrections in steps of 0.20 dex.  \label{Fig:Si_cor}}
\end{center}
\end{figure}

Results for the ionisation corrections for Fe and Si are shown in Figures \ref{Fig:Fe_cor} and \ref{Fig:Si_cor} as a function of \nhI and log U. 
As defined above, for a ratio $X^{+}/H^{0}$ one
must add $\epsilon$ to get the true abundance $X/H$. 
Fe appears unaffected except in the extreme case of low \nhI and high ionisation parameter U. As was noted in \citet{Des03} and \citet{Mei07, Mei08} in all cases 
where a constraint on the ionization parameter could be placed in the modeled sub-DLA systems, log U$\la-2.5$, indicating that the correction factors are 
at most $\sim-$0.2 dex, in the sense that the measured abundances may be slightly higher than the true abundances. With low ionisation parameters U, Fe$^+$/Si$^+$
is an accurate indicator of Fe/Si, however at higher ionisation parameters (log U $\ga$-2), Fe$^+$/Si$^+$ is higher than the true Fe/Si, mimicking the effects 
of differential and higher depletion of Fe than Si.

\begin{figure}
\begin{center}
\includegraphics[angle=90, width=\linewidth]{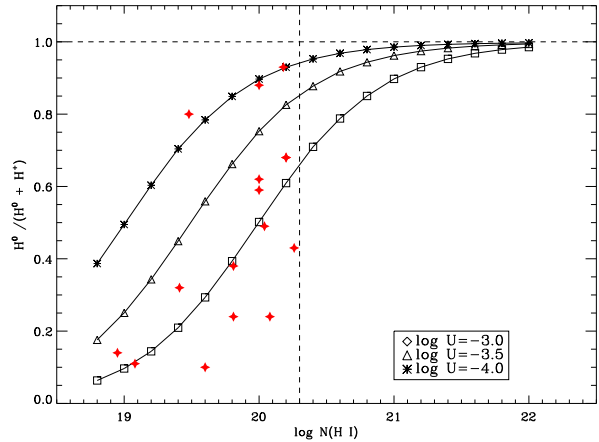}
\caption{ Results of the \textsc{cloudy} photoionisation models for the ionisation of hydrogen in these sub-DLA systems. The neutral fraction, $H^0/(H^0 + H^+)$ is plotted on the y axis, with 
log \nhI along the abscissa.  Also plotted as red stars are the ionisation fractions derived from \textsc{cloudy} models of each individual system. \label{Fig:HI_ions}}
\end{center}
\end{figure}

We have also investigated the fractional ionisation of hydrogen in these sub-DLA systems to determine the overall levels of ionisation which are needed to determine the 
contribution of sub-DLAs to the global metallicity. These simulations are more reliable than simulations of the ionisation of the metal 
species due to the direct calculability of the atomic properties of hydrogen such as the recombination rates, ionisation cross sections, and oscillator strengths. Again, 
we assumed a hot stellar spectrum with $T=30,000$ K, and the diffuse extragalactic background from AGN and starburst galaxies at $z\sim1$.  Constant density was assumed with $n=1$ atom cm$^{-3}$. 
Starting with log \nhI=18.8, the column density was increased by 0.2 dex in each subsequent iteration. 
Three separate ionisation parameters were used, with values of log U$=-4.0,-3.5$, and$ -3.0$.
These values were chosen because the sub-DLAs in this sample and \citet{Des03, Per07} all seem to have values of the ionisation parameter in or below this range.
Results of the simulations are shown in Figure \ref{Fig:HI_ions}. Overplotted in the figure are the values for the ionisation parameters
derived from individual \textsc{cloudy} models for the systems in this sample. 

In general, the sub-DLAs contain large fractions of ionized gas based upon these simulations, especially the systems with the lowest \nhI  while the DLAs are largely neutral in H.
This was also reported for higher $z$ systems by \citep{Per07}.
However, for a given \nhI a large range of ionisation fractions are seen, perhaps indicating that there are several types of environments from which 
the sub-DLAs are being drawn.

There are several unknowns however. The shape of the ionizing radiation field does affect the overall ionisation. The Haardt $\&$ Madau background spectrum is a
theoretical construct, and has not been accurately tested and as such is only one of many possibilities. The hot stellar spectrum used in these simulations, 
although certainly measurable in the Milky Way, may also not accurately reflect the true radiation field incident on these systems. 

%%%%%%%%%%%%%%%%%%%%%%%%%%%%%%%%%%%%%%%%%%%%%%%%%%%%%%%%%%%%%%%%%%%%%%%%%%%%%%%%
%%%%%%%%%%%%%%%%%%%%%%%%%%%%%%%%%%%%%%%%%%%%%%%%%%%%%%%%%%%%%%%%%%%%%%%%%%%%%%%%%

\section{Kinematics of Sub-DLAs \label{Sec:Kinematics}}

%%	Based on a sample of $\sim$ 53,000 star-forming galaxies at z$\sim$0.1 observed in imaging and spectroscopy in the SDSS, 
%%\citet{Trem04} discovered a mass-metallicity relationship for these galaxies. Specifically, they found a correlation between stellar
%%mass and metallicity that spans over 3 orders of magnitude in stellar mass and one order of magnitude in metallicity. 
%%Evidence has recently been provided for the possible existence of a mass metallicity relationship for DLA absorbers, 
%%assuming the velocity width of optically thin lines to be proportional to the mass \citep{Led06}. 
%%As a proxy for the stellar mass of these systems, which has been difficult to detect, the velocity width is used 
%%as an indicator of galaxy mass, as it potentially probes the depth of the gravitational potential well of the DLA systems. 
%%\citet{Zwaan08} however note that based on the 21 cm profiles of nearby galaxies, the mass does not correlate well with
%%the velocity width.

	Following the analysis of \citet{Led06}, we performed an analysis of the apparent optical depth (see e.g., \citealt{Sav96}) of these sub-DLA 
systems. The apparent optical depth is defined as

\begin{equation}
\tau(\lambda)_{a}=ln[I_{0}(\lambda)/I_{obs}(\lambda)]
\end{equation}

\noindent where I$_{0}(\lambda)$ is the continuum level, and I$_{obs}(\lambda)$ is the observed intensity.
Specifically, we integrated the AOD profile over the extent of the line, and give the velocity width 
($\Delta v_{90}$) as the value where the inner 90 percent of the absorption occurs. 
	
\begin{figure}
\begin{center}
\includegraphics[angle=90, width=\linewidth]{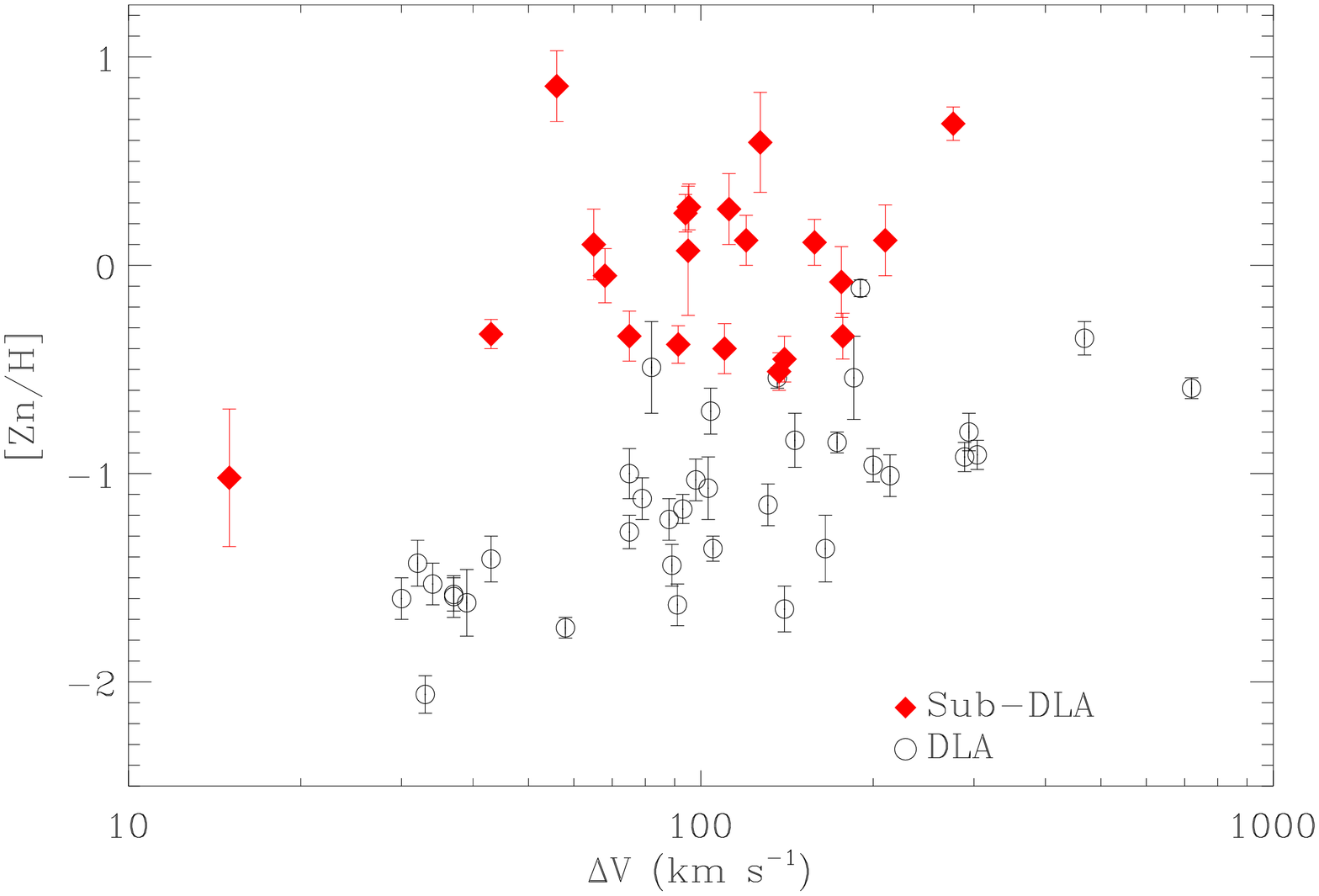}
\caption[ {$\Delta$V vs [Zn/H]}] { The metallicity as measured by the undepleted elements Zn and S vs the velocity width $\Delta v_{90}$ defined above. \label{Fig:ZntoHvsDV}}
\end{center}
\end{figure}

In Figure \ref{Fig:ZntoHvsDV} the metallicity vs $\Delta v_{90}$ for
the systems in \citet{Led06, Per06b, Per08, Pro06, Not08a, QRB08} as well as the systems from this work.   
In their analysis, \citet{Led06} used absorption features where 40 to 90 percent of the continuum level was absorbed by the line. We have tried to follow this 
strategy whenever possible for these systems. For most systems, the Fe II $\lambda$ 2374 line was used in the velocity width determination, as it is typically 
unsaturated in these systems, and also strong enough to include absorption from weaker features. For systems where the Fe II $\lambda$ 2374 feature was too weak, 
the Fe II $\lambda$ 2344 feature was typically used. From the \citet{Led06} sample, we have only included the Zn, 
and not the Si and S detections in Figure \ref{Fig:ZntoHvsDV}. 

%For the systems in our sample and \citet{Per06a,Per06b}, in addition to the Fe II $\lambda$ 2374 line,
%the  velocity width of the Mg II $\lambda$ 2796 or $\lambda$ 2803 line was also measured, 
%which is also plotted for each system in Figure \ref{Fig:ZntoHvsDV} with a line connecting the two values. 
%We have chosen these two lines for this investigation for several reasons. Firstly, with weaker features, the velocity 
%width becomes sensitive to the S/N.
%Lastly, the stronger Mg II $\lambda$ 2796, 2803 lines represent a maximum velocity width to the line, and also include the components at
%higher radial velocities that are often not seen in the intrinsically weaker features.

The non-parametric Wilcoxon Rank-Sum test was used to determine if the populations of sub-DLAs and DLAs were distinct in their mean velocity widths.
The test statistic $Z$ describing the significance
of the difference in the means of the two populations was
determined to be $Z=0.94$, with a probability of the sample populations having the same mean of 17 per cent. Similarly, a Kolmogorov–Smirnov
test shows that the populations of DLA and sub-DLA velocity widths having the same mean is $\sim$18 per cent. 
The non-parametric approach used here is more appropriate than a two sample
t-test, as we do not know the underlying distributions of the kinematics of these populations. 
A similar weak discrepancy was also seen by \citet{Zwaan08}, wherein the lower 
column density systems also had slightly larger $\Delta v_{90}$ values. 

As can be seen in Figure \ref{Fig:ZntoHvsDV}, the DLAs do show a fairly tight correlation between their metallicities and velocity widths. Interestingly, 
the sub-DLAs however do not show this trend. The Kendall's $\tau$ test was used for both the DLAs and sub-DLAs to test for a correlation between the metallicity and
velocity width. For the DLAs, a clear correlation is present with $\tau=0.59$ and a probability of no correlation $p<0.001$. Conversely, the sub-DLAs show no evidence of
correlation with $\tau=0.09$ and a probability of no correlation of $p\sim60$ percent. This however is possibly due to a lack of Zn detections at 
low [Zn/H].

%%%%%%%%%%%%%%%%%%%%%%%%%%%%%%%%%%%%%%%%%%%%%%%%%%%%%%%%%%%%%%%%%%%%%%%%%%%%%%%%%%%%%%%%%%%%%%%%%%%%%%%%%%%%%%%%%%%%%%%%%%%%%%%%%%%%%%%%%%%%%%%%%%%%%%%%%%%55
%%%%%%%%%%%%%%%%%%%%%%%%%%%%%%%%%%%%%%%%%%%%%%%%%%%%%%%%%%%%%%%%%%%%%%%%%%%%%%%%%%%%%%%%%%%%%%%%%%%%%%%%%%%%%%%%%%%%%%%%%%%%%%%%%%%%%%%%%%%%%%%%%%%%%%%%%%%

\section{\nhI-Weighted Mean Metallicity } \label{Sec:Weighted}			 																				
\subsection{[Fe/H] Metallicity Evolution} 

The \nhI-weighted mean metallicity given by: 
\begin{equation}
[\langle X/H \rangle]=log \langle (X/H) \rangle - log(X/H)_{\sun}
\end{equation}
\noindent where
\begin{equation}
\langle(X/H)\rangle=\frac{\sum_{i=1}^nN(X)_{i}}{\sum_{i=1}^nN(H)_{i}}
\end{equation}
\noindent with the errors estimated by the standard deviation, 
\begin{equation}
\sigma^2 = \left[\sum_{i=1}^n w_i([X/H]_i - [\langle X/H \rangle])^2 \right] / (n-1)
\end{equation}
\noindent and the weights proportional to \nhI as given by
\begin{equation}
w_i=\frac{N(H)_i}{\sum_{k=1}^n N(H)_k}
\end{equation}

\noindent has often been used as a quantitative way of estimating the amount of metal enrichment in the Universe in a given epoch \citep{Per03b, Kul05, Kul07}.
The \nhI-Weighted Mean Metallicity is thus a measure of the global or average chemical enrichment of the Universe in a given epoch or redshift interval. 				
																							
	As the lines of Fe are comparatively easier to detect and the cosmic abundance of Fe is substantially higher than that of the weaker Fe peak element Zn, Fe abundances for 
nearly every DLA and sub-DLA observed to date have been determined. There are also multiple lines of Fe II from $\sim$ 1200 to $\sim$2600 \ang, allowing the possibility for higher redshift 
systems to be studied, where the Zn II \lala 2026, 2062 lines would redshifted out of the visible range. On the other hand, as was previously mentioned, Fe is refractory and is typically
depleted onto dust grains. [Fe/H] is therefore a lower limit to the true abundance.

\begin{figure}
\begin{center}
\includegraphics[angle=0, width=\linewidth]{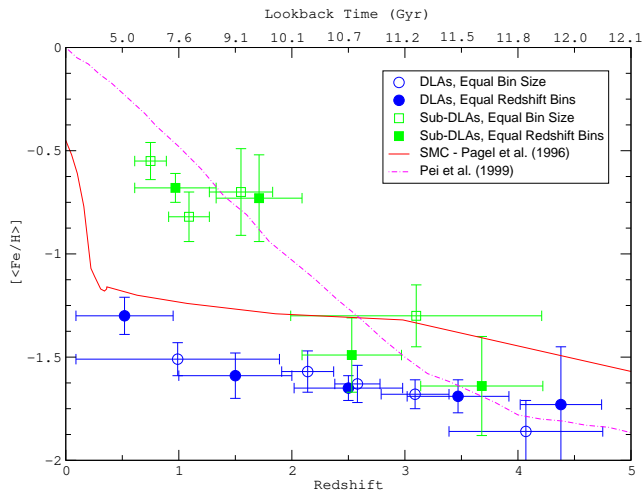}
\caption[Mean Fe Metallicity]{ The \nhI weighted mean metallicity for sub-DLAs (green squares) and DLAs (blue circles). Overlayed is the theoretical chemical evolution model from the 
 \citet{PFH99} model for the global metallicity (dashed line), and the chemical evolution model of the SMC from \citet{PT98} (solid line).
 The lookback time in Gyr is given on the upper x axis.  \label{Fig:FetoHvsZ}}
\end{center}
\end{figure}

	The \nhI-weighted mean metallicity based on Fe is shown in Figure \ref{Fig:FetoHvsZ}. Sub-DLAs have been plotted as green squares, and DLAs as blue circles. These data were binned 
into five sub-samples of constant redshift size ($\Delta z \sim 1$ and solid points) and into bins with roughly equal numbers of systems (open points). 
Independent of the binning type used, the sub-DLAs do appear to have a faster 
evolving metallicity, and are generally more metal rich in each redshift interval. Also shown in Figure \ref{Fig:FetoHvsZ} are the chemical evolution
models of \citet{PT98} for the SMC (solid red line), and \citet{PFH99} (dashed magenta line) for the global metallicity evolution of the Universe. Both types of 
models are sophisticated open box analytical models, with the models of \citet{PFH99} being averaged over several comoving Mpc to contain multiple galaxies.

\subsection{[Zn/H] Metallicity Evolution and Survival Analysis}

	Although far easier to detect in QSO absorbers, Fe suffers as a metallicity indicator due to dust depletion. The non-refractory elements of O, S, and Zn are 
much more reliable indicators of the true metallicity of the system. The O and S lines however lie far in the UV, at $\lambda$ $\sim$1300 and 
1250 \ang respectively and are inaccessible without space based
observations at $z\la2.0$. The Zn II \lala 2026, 2062 lines however are accessible with ground based observations for these systems. The nucleosynthetic
origins of Zn are still somewhat a matter of debate. With Z=30, Zn is not quite a true Fe peak element. The exact production site of Zn 
is still in question (see for instance the discussion in \citealt{Mish02}). 

The Kaplan-Meier estimator has become the standard non-parametric statistical analysis for dealing with
censored data \citep{KM58}. The Kaplan-Meier estimator works with any underlying distribution of the data, which is advantageous in the case 
of this data set for which the underlying distribution of metallicities at a given redshift or redshift interval is not known. 
We have used the \textsc{asurv} package provided by the Astrostatistics Center at Penn State University  (\verb http://astrostatistics.psu.edu ) for determining 
the summary statistics of the metallicities as measured by [Zn/H] for the DLAs and sub DLAs in this sample and the literature. The details of the
package are described in \citet{FN85}, for univariate data. 

\begin{figure}
\begin{center}
\includegraphics[angle=0, width=\linewidth]{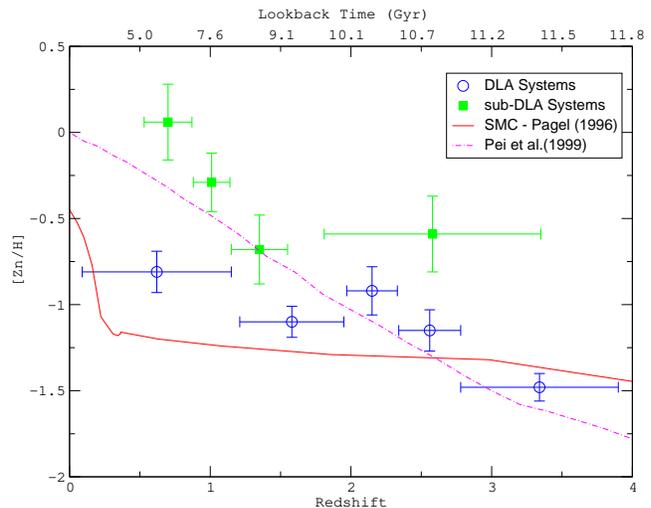}
\caption{ The metallicity via survival analysis taking into account upper limits for sub-DLAs (green squares) and DLAs (blue circles). 
Overlayed is the theoretical chemical evolution model from the 
\citet{PFH99} model for the global metallicity (dashed line), and the chemical evolution model of the SMC from \citet{PT98} (solid line). 
The lookback time in Gyr is given on the upper x axis \label{Fig:ZntoHvsZ}}
\end{center}
\end{figure}

The sub-DLA sample was separated into four bins with (roughly) 
equal numbers of systems in each bin, with the same done for the DLA systems from the literature along with the two DLAs from these observations. The DLAs were
separated into five bins due to the larger sample size. A plot of the metallicity as determined via this survival analysis is shown in Figure \ref{Fig:ZntoHvsZ}. 
The vertical error bars denote the estimate of the variance in each redshift
bin, and the horizontal bars denote the range of redshifts in each bin. 
Errors for the metallicity via survival analysis include only the statistical uncertainty and not
any measurement uncertainties. As can be seen in both Figures \ref{Fig:ZntoHvsZ} and \ref{Fig:FetoHvsZ}, the metallicity based on Fe and Zn
is faster evolving for the sub-DLA sample than the DLA sample. At $z\la1.25$, the sub-DLA metallicities are $\sim0.7$ dex higher than the DLA systems.

%%%%%%%%%%%%%%%%%%%%%%%%%%%%%%%%%%%%%%%%%%%%%%%%%%%%%%%%%%%%%%%%%%%%%%%%%%%%%%%%%%%%%%%%%%%%%%%%
%%%%%%%%%%%%%%%%%%%%%%%%%%%%%%%%%%%%%%%%%%%%%%%%%%%%%%%%%%%%%%%%%%%%%%%%%%%%%%%%%%%%%%%%%%%%%%%%

\section{Average Sub-DLA Spectrum } \label{Sec:Averaged}

\citet{York06} used a sample of $\sim$ 800 lower resolution ($\sim$2000) SDSS spectra to determine abundances and depletion
levels in QSO absorbers by averaging the spectra after shifting them to the
rest frame of the absorbers. Weaker absorptions, not seen or seen at lower
significance in the individual SDSS spectra, were clearly detected in this
average spectrum owing to its higher S/N ratio. Following in the spirit of that work, we have also averaged the spectra of these sub-DLAs.

\begin{figure}
\begin{center}
\includegraphics[angle=90, width=\linewidth]{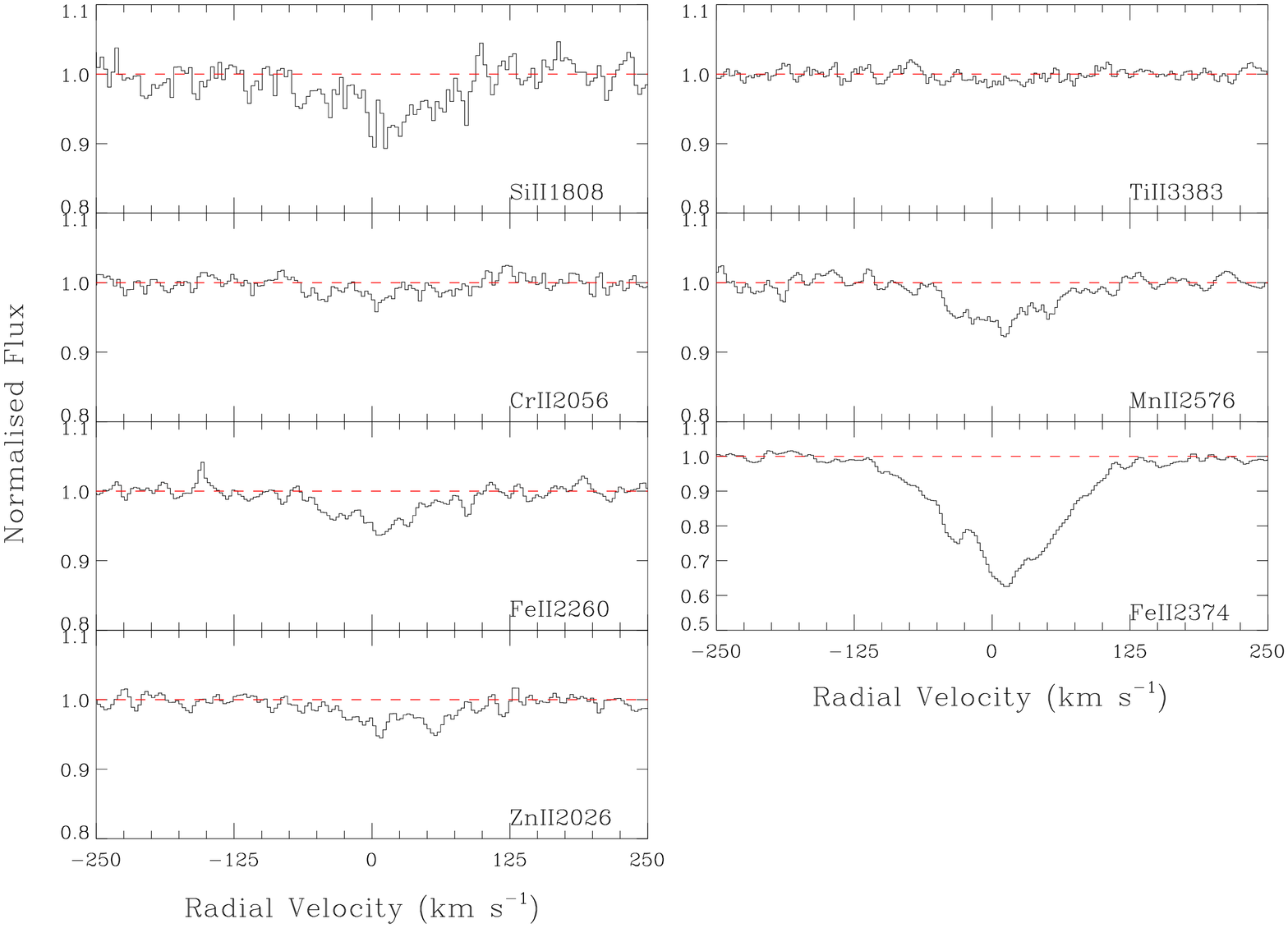}
\caption[Averaged Spectra] { The average spectra of the systems in this sample. The key lines of Si II, Fe II, Zn II, Cr II, Mn II are shown here. \label{Fig:AverageSpectra}}
\end{center}
\end{figure}

To determine average abundances of Cr, Fe, Mn,Si, and Ti of the absorbers in this sample, the spectra were shifted to rest wavelengths and averaged. 
Specifically, the Si II $\lambda$ 1808, Cr II $\lambda$ 2056, Fe II \lala 2260, 2374 and Zn II $\lambda$ 2026 lines were used. 
Any obvious non-associated features in the region around the lines of interest in the individual spectra were replaced by Gaussian noise with S/N similar to the uncontaminated regions.
For Q0251+1950, a systematic shift of $\sim$60 \kms was applied to the spectra as the 
redshift reported in \citet{Rao06} was slightly off. All other spectra were left unaltered. 
We show the averaged spectra of these transitions in Figure \ref{Fig:AverageSpectra}. 

Column densities for these elements were determined via the AOD method. The profile fitting method, although more robust and generally more 
accurate can not be used in this case, as the individual components are blended together in the averaged spectra. Rest frame equivalent widths were also determined, with the errors 
estimated by both photon noise and uncertainties in the continuum placement. 
Absolute  abundances were determined by averaging \nhI for the systems that were included in the averaged metal line spectrum, while excluding the systems which were not used in the average.  
Table \ref{Tab:AvgSpectra} gives \w, the measured column density and corresponding abundance. Interestingly, the metallicity based on Zn from this averaged spectrum
matches nearly perfectly with the \nhI-weighted mean metallicity from \S\ref{Sec:Weighted}, validating the technique. 

\setlength{\tabcolsep}{5pt}
\begin{table}
\caption[Properies of Averaged sub-DLA Lines]{Equivalent widths, abundances and column densities of the averaged spectra. \label{Tab:AvgSpectra} }
\begin{center}
\begin{tabular}{lllll}
\hline
\hline	
Transition	&  $\#$ Systems 	   	& 	\w		& 		log N			&	[X/H]				\\
		&				&	m\AA		&		cm$^{-2}$		&	Dex			\\
\hline
Si II 1808	&	15		& 	44$\pm$9	& 		14.87$\pm$0.08		&	$-0.52\pm0.08$			\\
Ti II 3383	&	15		&	12$\pm$4	&		11.52$\pm$0.17		&	$-1.14\pm0.17$			\\
Cr II 2056	&	29		& 	15$\pm$4	&		12.59$\pm$0.12		&	$-0.91\pm0.12$			\\
Mn II 2576	&	25		& 	51$\pm$8	&		12.39$\pm$0.06		&	$-0.98\pm0.06$			\\
Fe II 2260	&	29		&	37$\pm$5	&		14.53$\pm$0.06		&	$-0.79\pm0.06$			\\
Fe II 2374	&	28		&	327$\pm$7	&		14.38$\pm$0.01		&	$-0.93\pm0.03$			\\
Zn II 2026	&	31		&	28$\pm$5	&		12.10$\pm$0.08$^a$	&	$-0.38\pm0.08^a$		\\
\hline

\end{tabular}
\begin{minipage}{\linewidth}
$^a$Based on the average column density determined with and without the feature at $\sim+60$ \kms, which could be due to a contribution from Mg I 2026.
\end{minipage}
\end{center}

\end{table}

\begin{figure}
\begin{center}
\includegraphics[angle=90, width=\linewidth]{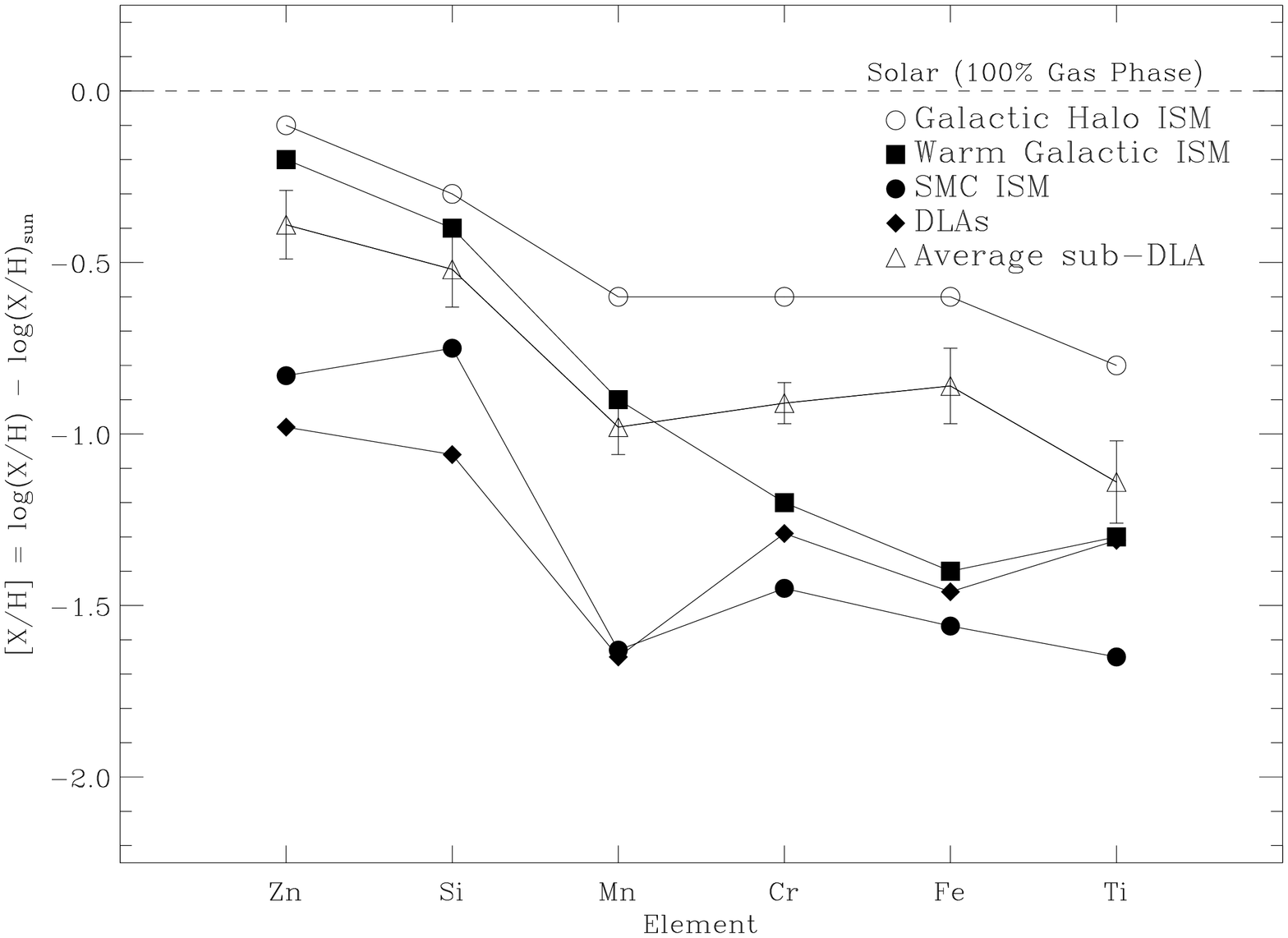}
\caption[Averaged Absolute Abundances]{ The depletion pattern or 
abundances derived from the averaged sub-DLA spectra of the systems in this sample. Overplotted are the abundances of the Milky Way warm neutra ISM, Milky Way halo,  
ISM abundances in the SMC, and DLA abundances.  \label{Fig:AvgDepletion}} 
\includegraphics[angle=90, width=\linewidth]{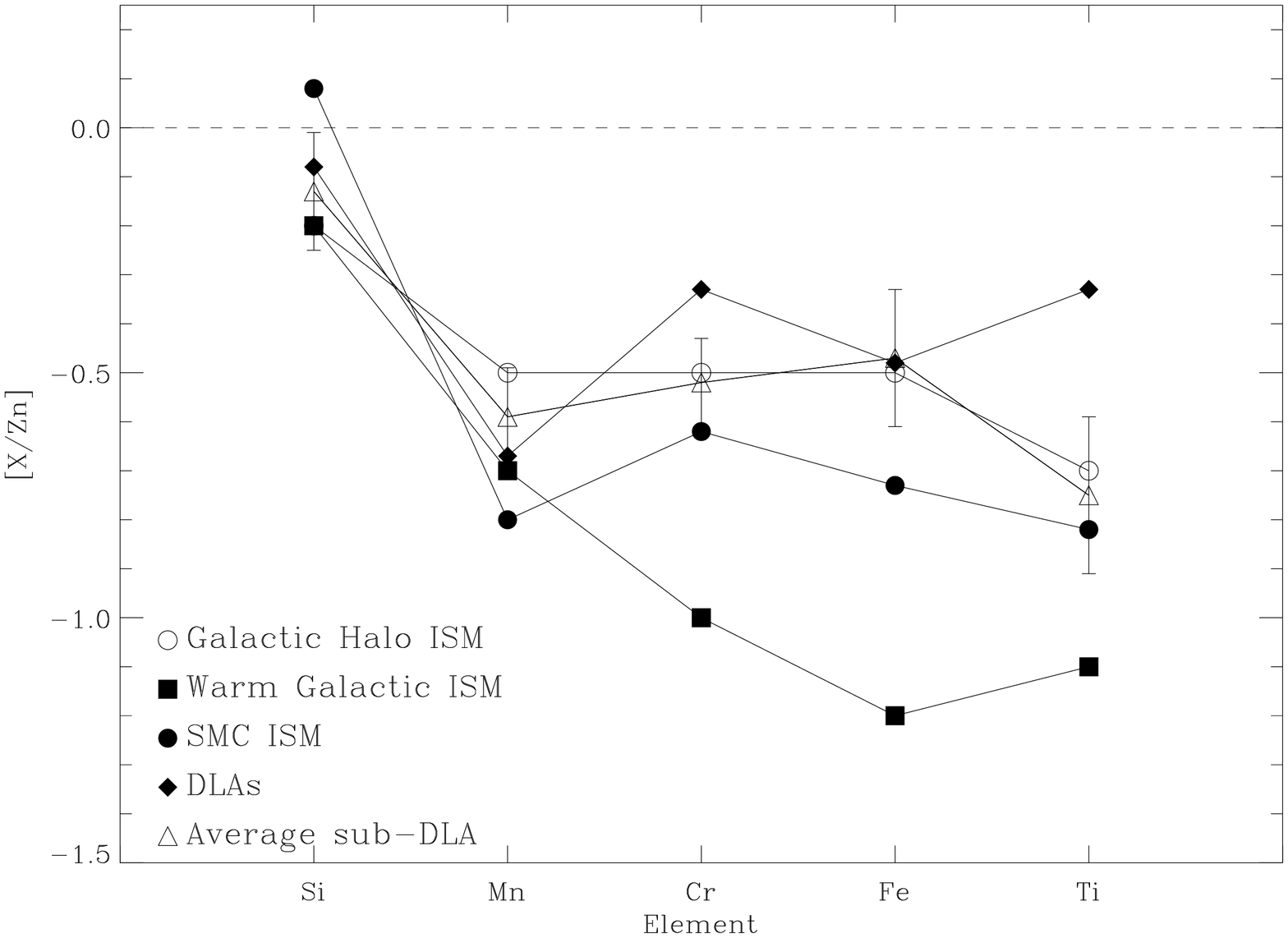}
\caption[Averaged Relative abundances]{ The abundances relative to Zn derived from the averaged sub-DLA spectra of the systems in this sample. 
Overplotted are the abundances of the Milky Way warm ISM, Milky Way halo,  
ISM abundances in the SMC, and DLA abundances. \label{Fig:AvgRelative}} 
\end{center}
\end{figure}

	The abundance patterns seen in the ISM of the Milky Way, SMC, and DLAs are shown in Figure \ref{Fig:AvgDepletion}. Abundances from the ISM of the SMC are from \citet{Wel97} with 
abundances from the DLAs from \citep{PW02}.
Several points come to light based on the depletion patterns and abundances of these samples. The ISM Si abundance is nearly identical in all three cases.  Secondly, the Zn abundance 
of the averaged sub-DLA spectra lies between the SMC and warm Galactic ISM value. Lastly, the Fe and Cr abundances in the averaged sub-DLA spectra are somewhat higher
($\sim+0.5$ and  $\sim+0.3$ dex for Fe and Cr respectively) than what is observed in the warm neutral ISM of the Milky Way and SMC. 

	As the abundance patterns seen in the ISM are closely related to the past star formation rate and history of the galaxy, does the average abundance pattern in sub-DLAs 
mimic the patterns seen in the Milky Way or ISM, or neither? The SMC and Milky Way have had very different star formation histories, and thus have 
different elemental abundances. The star formation history of the SMC has recently been investigated by \citet{HZ04} based on deep photometric surveys of the entire SMC. 
They find that most of the stars in the SMC formed $\ga$8.5 Gyr ago (corresponding to redshift $z \ga 1.2$) with a period of quiescence lasting $\sim$3 Gyr, followed by more recent 
bursts in star formation. Tidal interactions with the Milky Way likely play a roll in causing star formation bursts in the SMC \citep{HZ04}. The Milky Way has 
also gone through bursts of star formation, 
creating the halo and thick disk. The Milky Way however does not appear to have undergone the same long duration periods of quiescence that the SMC underwent, but instead has 
seemed to continuously form stars \citep{Gil01, RP00}. Sub-DLAs may trace a population of galaxies dissimilar to massive spirals (i.e. the Milky Way) or lower mass 
irregulars (i.e. the SMC) and thus may have different abundance patterns.

    Based upon the average abundance pattern in these sub-DLA systems, is appears as though these systems are more closely related to the Milky Way halo type interstellar medium than the 
warm neutral ISM of the Milky Way. The Zn metallicity in sub-DLAs, [Zn/H], as measured in the averaged spectra lies between the values seen in the warm neutral ISM and SMC.
 [Zn/H] is $\sim$0.2 dex less than that of the warm  Milky Way ISM and 0.3 dex higher than the SMC. This however reflects [Zn/H] at $z \sim 1.1$, the average redshift of these systems. 
This metallicity is well within the range of values at this redshift from the chemical evolution models mentioned above. The abundance and depletion pattern of the sub-DLAs does 
very closely resemble that of the halo of the Milky Way, as can be seen in Figure \ref{Fig:AvgRelative}.

%%%%%%%%%%%%%%%%%%%%%%%%%%%%%%%%%%%%%%%%%%%%%%%%%%%%%%%%%%%%%%%%%%%%%%%%%%%%%%%%%%%%%%%%%%%%%%%%
%%%%%%%%%%%%%%%%%%%%%%%%%%%%%%%%%%%%%%%%%%%%%%%%%%%%%%%%%%%%%%%%%%%%%%%%%%%%%%%%%%%%%%%%%%%%%%%%

\section{The Contribution of sub-DLAs to the Metal Budget of the Universe } \label{Sec:Global}

	It has now been well established that the DLA systems observed to date are metal poor at all redshifts, and show weak evolution with redshift \citep{Kul05, PW02, Mei06}. 
Including the contributions from star forming Lyman-break galaxies at $z\sim2$, DLAs, and the \Lya forest the observations account
for only $\sim60\%$ of the amount of metals that are expected from the cosmic star formation history \citep{Bou08}. Here, we discuss the contribution of the less studied sub-DLA systems 
to the metal budget of the Universe. For a previous estimation of the comoving metal density in sub-DLAs based on the preliminary results of this work and the
literature sample of sub-DLAs, see \citet{Kul07}.

We can estimate the comoving metal density in DLAs and sub-DLAs based on the column density distribution function, metallicities, and in the case of sub-DLAs the ionisation fraction. 
As was discussed previously, and can be seen in Figure \ref{Fig:HI_ions}, the ionisation fraction in the DLA systems with log \nhI$>$20.3 is negligible, and they can be considered completely neutral
to  good approximation in the H I phase. See however \citet{WP00} for a discussion on the more highly ionized gas in DLAs.

	The DLAs are well sampled, especially at $z > 1.5$, and have a mean metallicity of [Zn/H]$\sim$-1.0 from a robust sample of $\sim$100 systems \citep{PW02, Kul05, Mei06}. Using the 
solar metallicity by mass as determined by \citet{Lodd03} of $\bar{Z}_{\sun}$=0.0177, the DLAs thus have a metallicity by mass of $\bar{Z}_{DLA}$=0.00177. The fraction of mass in metal ions 
in DLA systems compared to the critical density of the Universe is thus

\begin{equation}
\Omega_{met}^{DLA} = \frac{\rho_{met}}{\rho_{crit}} = \Omega_{H I}^{DLA} \cdot \bar{Z}_{DLA}
\label{Eq:OmegaDLA}
\end{equation}

\noindent \citet{Per05} measured $\Omega_{HI}^{DLA}$ at $z\sim2.5$, where the bulk of DLA metallicities are determined, to be $\Omega_{HI}^{DLA} \sim 0.85\times10^{-3}$. 
This then gives the comoving mass density of metals in DLAs to be $\Omega_{met}^{DLA}\sim2\times10^{5}$ M$_{\sun}$ Mpc$^{-3}$ (see also \citealt{Kul07, Pro06}). 

For sub-DLA systems, the ionisation fraction of the gas must be taken into account to determine the comoving mass density of metals contained in these systems. 
If we let $y(N_{\rm HI})$ be the neutral fraction of gas in sub-DLAs (which is a function of the neutral hydrogen column density as can be seen in
Figure \ref{Fig:HI_ions}), $f(N_{\rm HI})$ the column density distribution in the sub-DLA regime, and the average metallicity $\bar{Z}_{subDLA}$
we have then, 

\begin{equation}
\Omega^{\rm{subDLA}}_m = \frac{\mu H_0 m_H}{c \rho_{crit}} \int \frac{1}{y} \epsilon(Z) f(N_{\rm HI}) \bar{Z}_{\rm{sub}}N_{\rm HI}  dN_{\rm HI}
\label{Eq:Omega_subDLA}
\end{equation}	

\noindent where $\epsilon (Z)$ is the ionisation correction term, which as can be seen from Figures \ref{Fig:Fe_cor} and \ref{Fig:Si_cor} is a function of \nhI and the integral runs over the
sub-DLA column density range of $19.0 <$ log \nhI $< 20.3$. We have used the ionization correction determined for Fe for the purpose of these calculations due to the 
better quality of atomic data for Fe.   

Unfortunately, at lower redshifts the comoving gas density in sub-DLA systems is not well known, owing to the lack of space based UV surveys for H I at low $z$. Recently however, 
\citet{Omear07} and also \citet{Per05} determined the column density distribution function (CDDF) and comoving mass density of gas in sub-DLAs from observations with the Keck ESI,
Magellan II MIKE spectrograph, and the VLT Kueyen UVES spectrograph. Combining the observations from both teams, we find that the CDDF is well fit by a power law 
($f(N_{\rm HI})=\beta N_{\rm HI}^{\alpha}$ with $\beta=10^{4.84}$ and $\alpha=-1.30$ in the sub-DLA regime of 19.0 $<$ log \nhI $<$ 20.3.  

Survival analysis provides an estimate of the \nhI-weighted mean metallicities given in Table \ref{Tab:subDLA_Comoving}.
Using these values for the column density distribution 
function and mean metallicities, Equation \ref{Eq:Omega_subDLA} can be integrated. As can be seen in Figures \ref{Fig:FetoHvsZ} and \ref{Fig:ZntoHvsZ},  
the sub-DLA and DLA populations 
are dissimilar in their respective metallicities, 
sub-DLAs having a significantly higher \nhI-weighted mean metallicity of $\langle$[Zn/H]$_{subDLA}\rangle\sim-0.08$ 
and $\langle$[Zn/H]$_{DLA}\rangle\sim-0.8$ at $z\la1$.
These calculations were carried out using three ionisation parameters and thus three separate levels of ionisation for these systems.
We have used the results from the photoionisation models for Fe for the correction term $\epsilon (N_{\rm HI})_{\rm Fe II}$,  when evaluating  
Equation \ref{Eq:Omega_subDLA}. 

Results for the calculations of the 
comoving mass density of metals in the sub-DLAs are given in Table \ref{Tab:subDLA_Comoving} 
The sub-DLAs thus appear to contain several times the comoving mass density of metals that the DLA systems contain at $z<1$, and at least equal amounts at all redshifts. 
We stress however that the comoving mass density of H, 
$\Omega_{HI}^{DLA}$, is not well known at low $z$ and warrants further investigation.

The more highly ionized phase of sub-DLAs has also been investigated as a possible location of the missing metals. At low density and temperatures 
$T\sim10^{6}$ K, the cooling time of the gas is roughly the age of the universe, so metal atoms can be essentially stuck in this hot phase. 
\citet{Fox07} estimates that the metal content of the hot plasma in sub-DLAs at $z\sim2.5$ is roughly equal to the what we have determined here for the 
lower temperature gas in which Zn II arises. 

\begin{table}
\begin{center}
\begin{tabular}{llll}
\hline
\hline	
Redhsift  	&	Metallicity	  	& 	$\Omega_Z^{a} $		&	$\rho_{Z}^{a}$			\\	
Range		&	Dex			&				&	 M$_{\sun}$ Mpc$^{-3}$		\\
\hline
0.53 - 0.87	&	$+0.06\pm0.22$		&(7.8-31.3)$\times10^{-6}$	&	(10.6-42.5)$\times10^{5}$	\\
0.89 - 1.14	&	$-0.29\pm0.17$		&(3.5-13.9)$\times10^{-6}$	&	(4.7-18.9)$\times10^{5}$		\\
1.15 - 1.55	&	$-0.68\pm0.20$		&(1.4-5.7)$\times10^{-6}$	&	(1.9-7.8)$\times10^{5}$		\\
1.77 - 3.39	&	$-0.59\pm0.22$		&(1.8-7.0)$\times10^{-6}$	&	(2.5-9.5)$\times10^{5}$		\\					
\hline
\end{tabular}
\end{center}

\caption[$\Omega_{met}^{subDLA}$]{The comoving mass density of metals, $\Omega_{met}^{subDLA}$ and $\rho_{Z}$  in sub-DLAs. \\
	$^{a}$ Calculated based on the $z\sim2$ column density distribution discussed in the text. 
 \label{Tab:subDLA_Comoving} }
\end{table}

%%%%%%%%%%%%%%%%%%%%%%%%%%%%%%%%%%%%%%%%%%%%%%%%%%%%%%%%%%%%%%%%%%%%%%%%%%%%%%%%%%%%%%%%%%%%%%%%
%%%%%%%%%%%%%%%%%%%%%%%%%%%%%%%%%%%%%%%%%%%%%%%%%%%%%%%%%%%%%%%%%%%%%%%%%%%%%%%%%%%%%%%%%%%%%%%%

\section{Discussion and Conclusions} \label{Sec:Discussion}

We have compiled all of the known Zn measurements in DLAs and sub-DLAs, and have examined log \nhI vs. [Zn/H]. As has been reported in the past, the observed trend towards increasing metallicity 
with decreasing \nhI is again seen with the inclusion of these new data \citep{Boi98, Kh07, Mei06} . We have used survival analysis to calculate a trend line to these data including the upper limits, which is determined to
be of the form [Zn/H]$=-0.58$\nhI + $10.9110$. Without inclusion of the upper limits in the trend line, we determined [Zn/H]$=-0.78$\nhI$+15.27$. Both Spearman and Kendall $\tau$ tests independently 
determined that the probability of no correlation is $p<0.001$, with the inclusion of the upper limits. A similar trend was seen in \citet{York06} using composite spectra of
a large number of SDSS QSO absorbers, and in \citet{Kh07} using the literature data as well. 

\begin{figure}
\begin{center}
\includegraphics[angle=0, width=\linewidth]{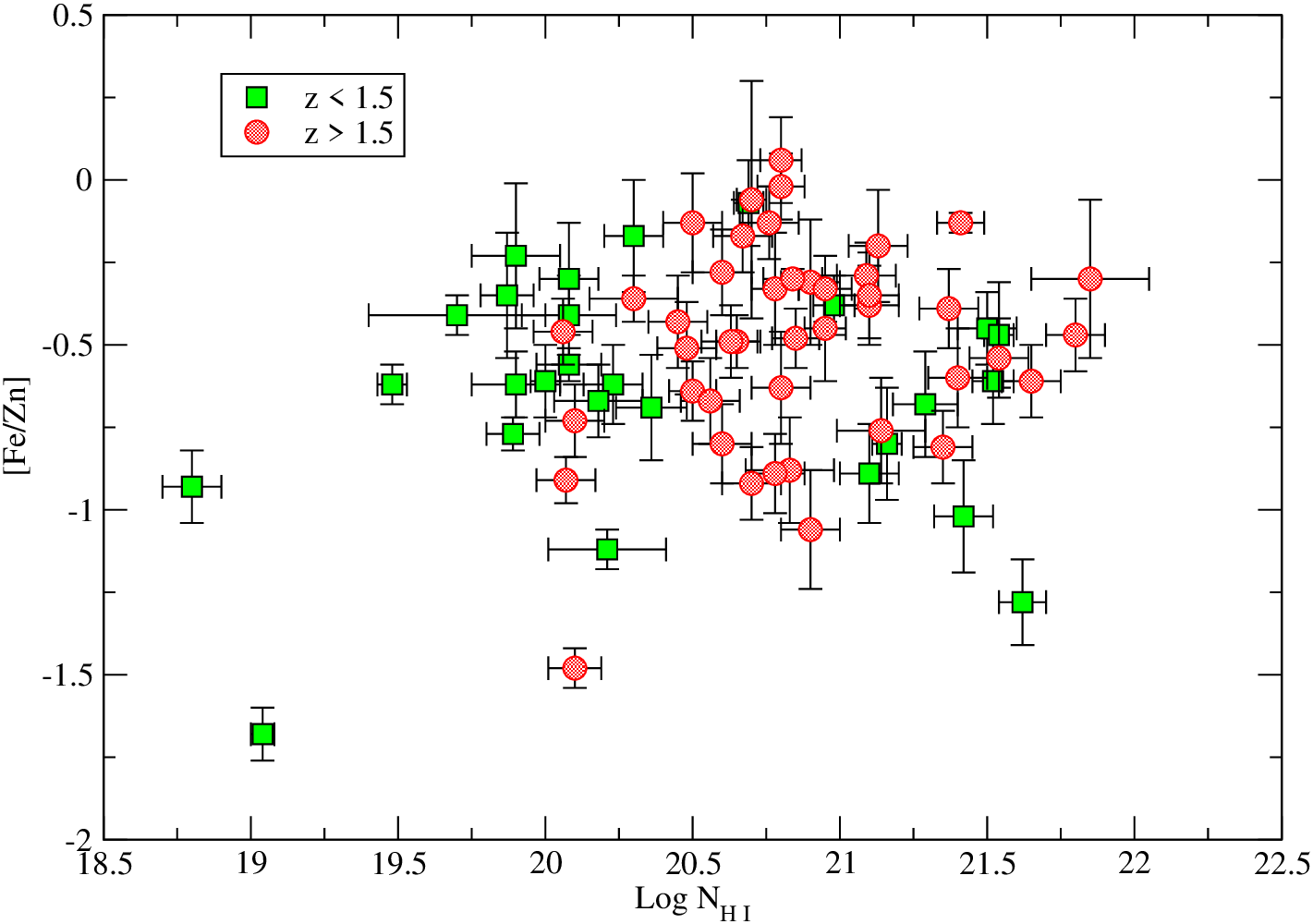}
\caption{  [Fe/Zn] vs. Log \nhI for QSO absorbers with both Fe and Zn detections from this work, as well as systems from \citet{Kul07, Des03, Pro06, QRB08, Not08a}. \label{Fig:FetoZnvsH} }
\vspace{0.8cm}
\includegraphics[angle=0, width=\linewidth]{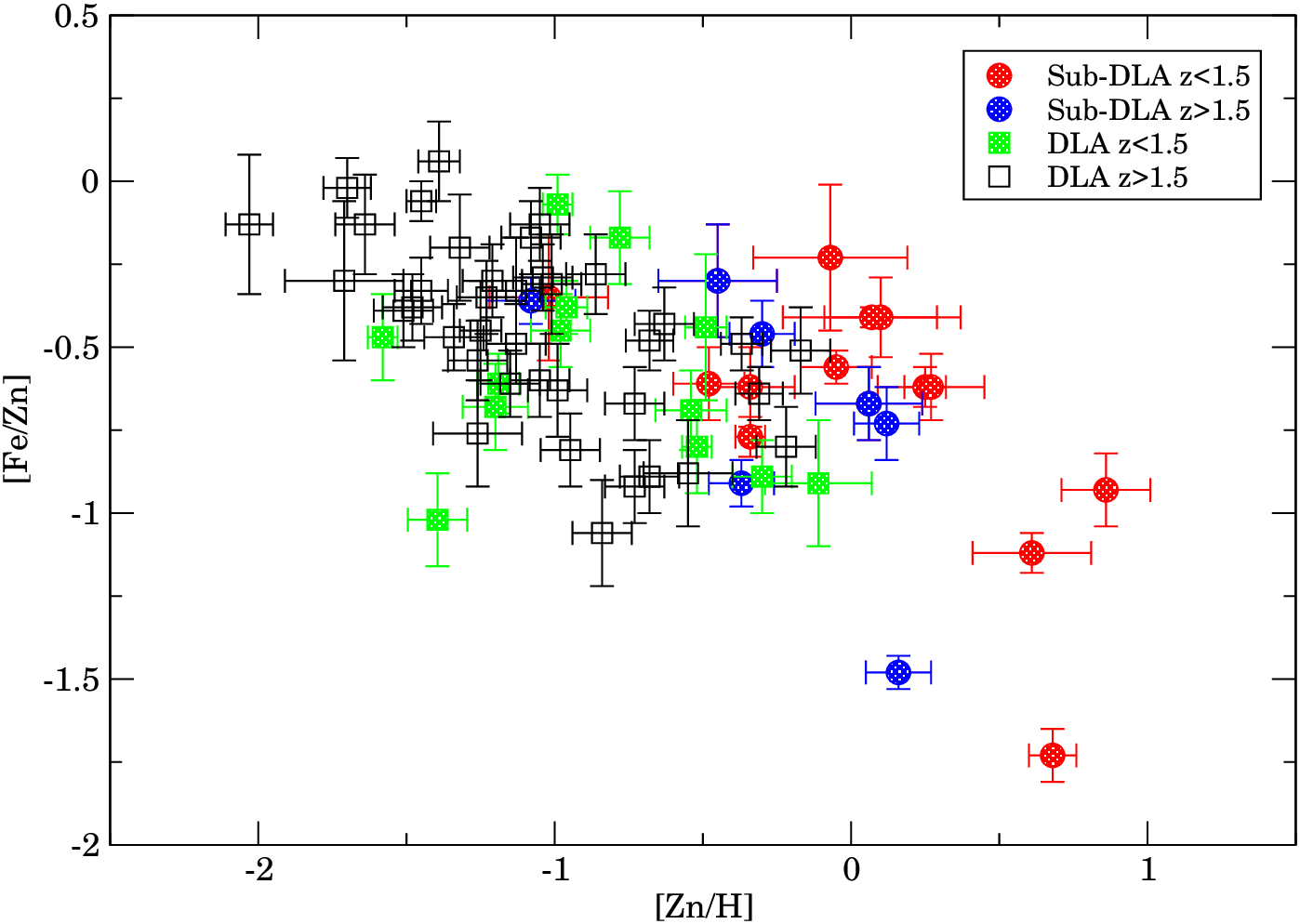}
\caption{  [Fe/Zn] vs. [Zn/H] for QSO absorbers with both Fe and Zn detections from this work, as well as systems from \citet{Kul07, Des03, Pro06, QRB08, Not08b}. \label{Fig:FetoZnvsMet} }
\end{center}
\end{figure}

Interestingly, the amount of dust as estimated by [Fe/Zn] seems independent of
\nhI in QSO absorbers over $\sim$3 decades in \nhI as is shown in Figure \ref{Fig:FetoZnvsH}. Spearman and Kendall correlation tests
each reveal a probability of no correlation of $p\sim84$ percent. The dust depletion on the other hand is known to correlate 
with the metallicity of the system as shown in Figure \ref{Fig:FetoZnvsMet} where we show [Fe/Zn] vs [Zn/H] for 
QSO absorbers with both Fe and Zn detections \citep{Kul07, Des03, QRB08, Not08a}. Spearman and Kendall correlation tests show that the probability of no
correlation between the metallicity and depletion as estimated by [Fe/Zn] to be $p<0.001$. 

\begin{figure}
\begin{center}
\includegraphics[angle=0, width=\linewidth]{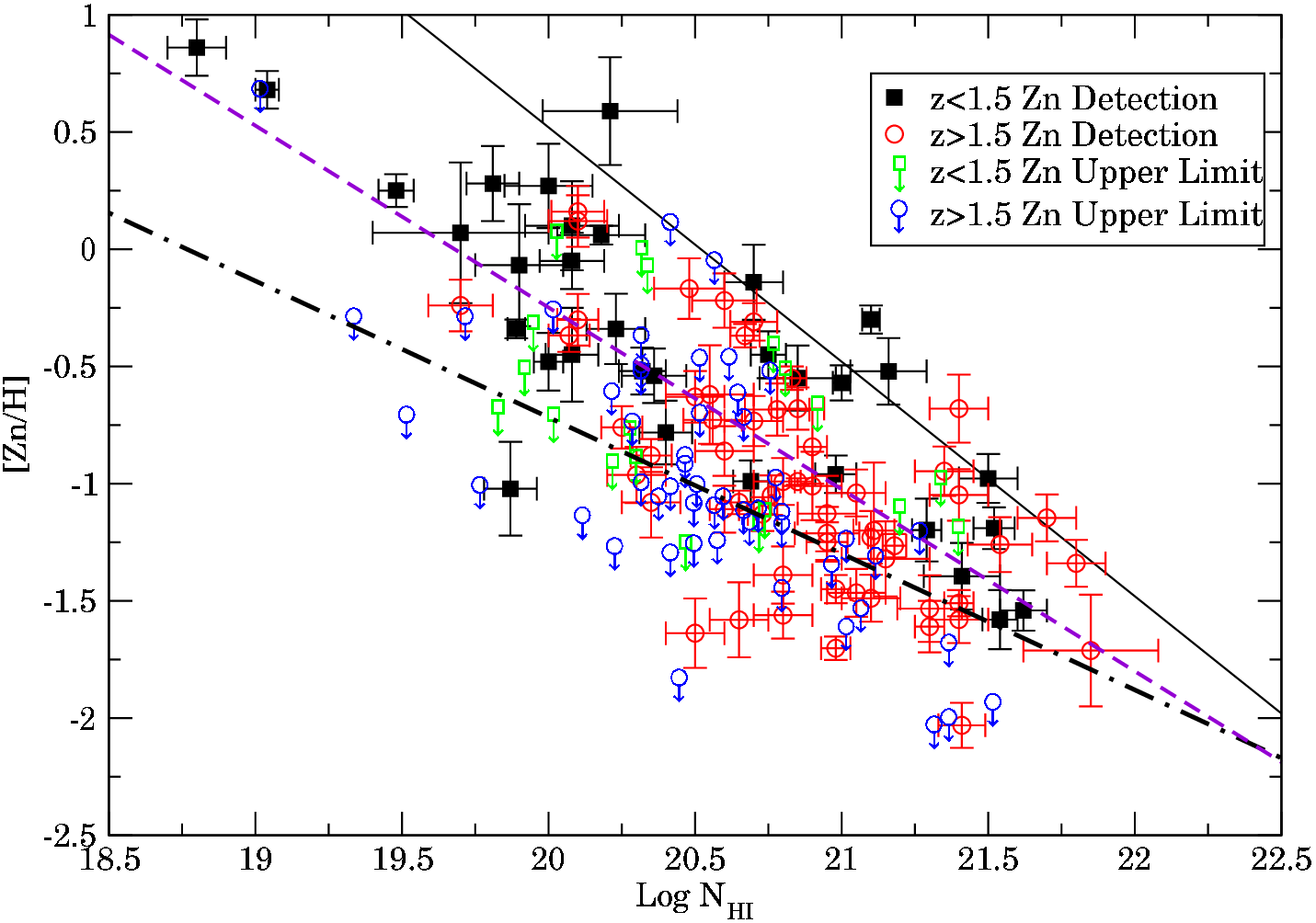}
\caption{ Log \nhI  vs. [Zn/H] for all the systems given in \citet{Kul07}, systems from this sample,  and systems from 
\citet{QRB08, Not08a, Pro06, Des03}. The solid line is the ``Obscuration Threshold'' from \citet{Boi98}, while the dashed line is the linear fit to these data ignoring 
the [Zn/H] upper limits, and the dashed-dotted line is the linear fit while including the upper limits. \label{Fig:ZntoHvsHI}}
\end{center}
\end{figure}

The observed trend of increasing metallicity with decreasing \nhI seen in Fig. \ref{Fig:ZntoHvsHI} contradicts measurements in the local ISM of the Milky Way and
the Magellanic Clouds, where different lines of sight show roughly solar metallicity gas although the sightlines may contain a range of \nhI values. 
Abundance gradients could not produce the effect as is seen; if the majority of DLAs are produced in the outskirts of the disks of massive spirals where 
the abundances would be lower due to the gradient, a simultaneous gradient of decreasing \nhI with decreasing galactic radius which is not observed 
would be necessary to explain the lower \nhI systems with higher abundances. 

In all likelihood, the sightlines through DLAs and sub-DLAs pass through gas rich galaxies of a wide range of morphologies and masses. The fact that the ionization fractions 
in sub-DLAs from the models above as well as in \citet{Per07} show a range of values for a given \nhI, as well as the fact that not all sub-DLAs are metal rich
suggest that this is indeed the case. There are however several lines of evidence to suggest that DLAs are mainly sampling low mass gas rich dwarf galaxies such as the SMC:

\begin{itemize} 
\item The abundances in DLA systems are low ([Zn/H]$\sim-0.8$) and slowly evolving, which is consistent with
the observations and chemical evolution models of local dwarf galaxies. The \nhI-weighted mean metallicity of DLAs does not track the global chemical evolution 
models and is essentially flat for $1<z<3$, which is predicted by \citet{PT98} for the SMC. 

\item The molecular fraction of gas in DLAs is small \citet{Led03}, consistent with CO surveys of local dwarf galaxies \citep{Leroy05}. The 
molecular fraction in dwarf galaxies is substantially smaller than in more massive systems \citep{Leroy05}, 
consistent with FUV stellar observations in the Magellanic Clouds (e.g. \citealt{Tum02}).
Although H$_{2}$ was detected in most of the SMC sight lines from \citet{Tum02}, the molecular content of the SMC was 
likely much lower in the past as there was a long quiescent period in the star formation history of the SMC. 
The molecular content of sub-DLAs has not been investigated as thoroughly, however molecule rich sub-DLAs have been observed \citep{QRB08, Not08a}.

\item Followup imaging studies of DLAs show that the majority of the galaxies associated with the DLA absorption, 
when detected have $L<L_{\star}$ \citep{LeBrun97, Rao03, Kul06}. The low mass local dwarf galaxies
observed in \citet{Geha06} were determined to have total baryonic masses actually dominated by the gas, and not the stellar content, with gas fractions as high
as $f_{gas}=95\%$ and $\langle f_{gas} \rangle \sim 0.6$. This could explain the low luminosities observed in DLA galaxies, and the low detection rate in followup imaging. 
\end{itemize}

This still raises the question as to \emph{why} we do not see metal rich DLAs and do see more metal rich systems at lower \nhI, 
and the observed relationship between [Zn/H] and \nhI. One proposed explanation is that there is a selection effect causing the
higher metallicity DLAs to be systematically undersampled due to dust obscuring the background QSO to magnitudes fainter than what 
is observable with the current classes of instrumentation \citep{Lau96, Boi98, VP05}. The amounts of dust in QSO absorbers as estimated by ratios of 
volatile and refractory elements such as [Cr/Zn] or [Fe/Zn] is strongly correlated with the systems metallicity \citep{Mei06} the higher 
metallicity systems thus contain more dust and could be obscuring the background QSO. 

\begin{figure}
\begin{center}
\includegraphics[angle=90, width=\linewidth]{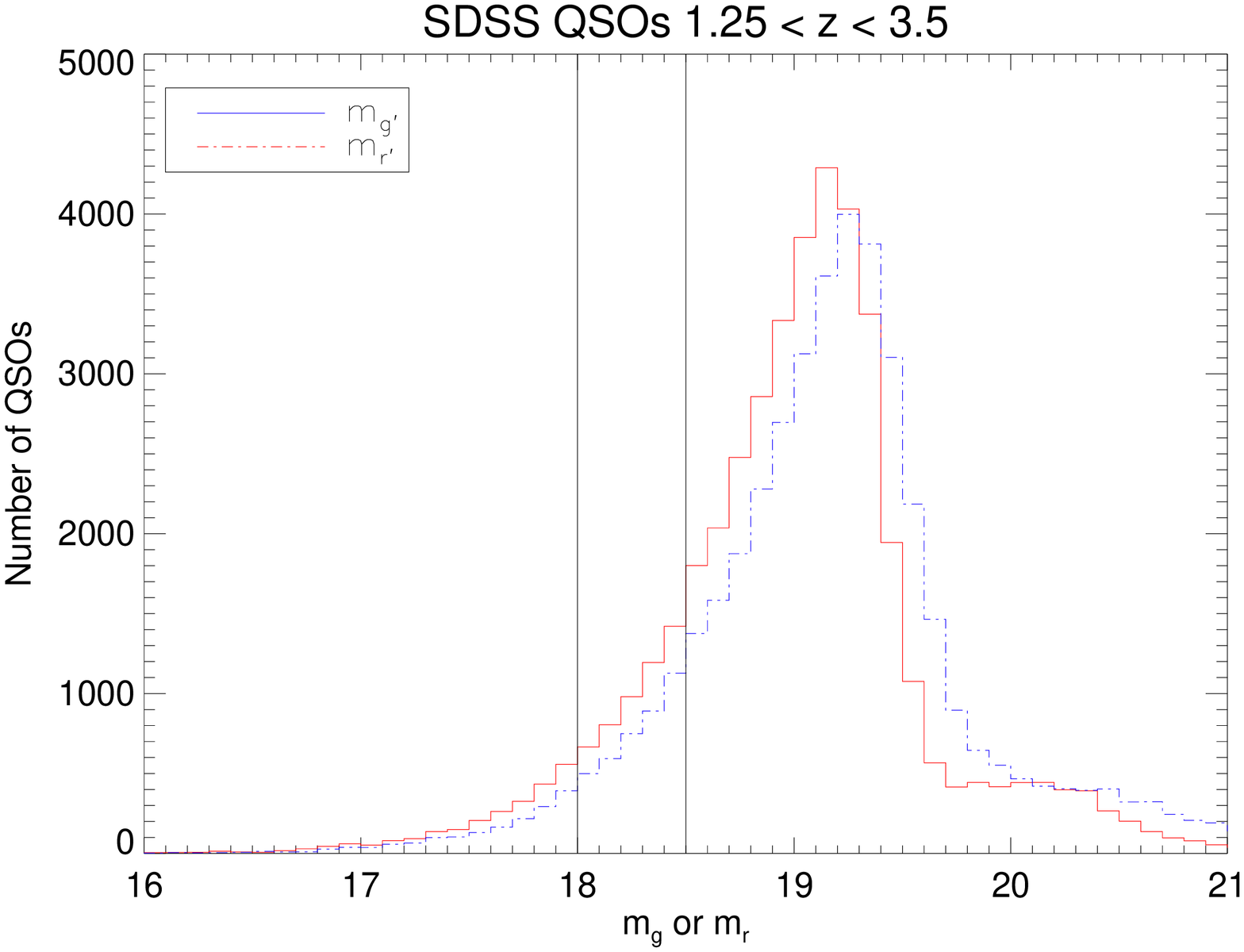}
\caption{The distribution of SDSS g$^{\prime}$ and r$^{\prime}$ magnitudes of QSOs with $1.25<z<3.5$. The vertical lines at $m=$18.0 and $m=$18.5
represent the approximate magnitude limitations of the current instrumentation with and without an extinction effect respectively.  \label{Fig:SDSS_QSOs} }
\end{center}
\end{figure}

The amount of extinction need not be 
very large to exclude a large number of background QSOs from being observed. In fact, as can be seen in Figure \ref{Fig:SDSS_QSOs}, 
only $\sim4$ percent of SDSS DR7 QSOs have m$_{g\prime}<18.0$ at $1.25<z_{em}<3.5$, so even
with a modest 0.5 magnitudes of extinction (less than 50 percent of the estimated extinction amounts in \citet{VP05} for a solar metallicity DLA) we would be left with only
4 percent of QSOs bright enough to  observe. As we are currently magnitude limited at $m_{g}\la18.5$ with current instrumentation,
a large number of DLAs could be lying unobserved in front of fainter QSOs. 

If this is indeed true, then one would expect that the \emph{observed} DLAs would have properties more similar to 
dwarf or SMC type galaxies. The DLAs do seem to exhibit a more gradual chemical evolution similar to the SMC, as seen in
Figures \ref{Fig:FetoHvsZ} and \ref{Fig:ZntoHvsZ}, and the molecular content along with the low luminosities of DLA host galaxies observed in 
imaging support the claim, as highlighted in the list above. With lower \nhI and thus likely lower levels of extinction, the sub-DLAs are less affected by
the dust extinction bias than the more H I rich traditional DLAs \citep{Lau96}. This could possibly explain why the mean abundances in these sub-DLA systems are
significantly higher than their DLA counterparts. An alternative explanation has been put forward by \citet{Kh07}, who suggest that based on the observed
mass metallicity relationship of \citet{Trem04} these higher metallicity sub-DLAs may arise in more massive spiral galaxies while the lower metallicity DLAs would be more similar to 
gas rich dwarfs.

More work still is needed in several areas though. 
More space based observations are necessary to constrain the low redshift column density distribution of sub-DLA systems, and hence $\Omega_{\rm HI}^{subDLA}$ which is currently 
poorly understood. If $\Omega_{\rm HI}^{subDLA}$ has varied from $z\sim2.5$ to $z\sim1.0$, the contribution of these sub-DLAs to the metal budget will
alter accordingly. 

The Lyman limit systems with 17.2$<$log \nhI$\la$19.0 are yet another possible reservoir of metals that have been hitherto  
understudied, although progress is being made (see for instance \citealt{Char03, Mis08}). 
Several difficulties arise in these systems as \nhI cannot be constrained due to the lack of damping wings in the \Lya profile in this 
column density range, and with such low neutral hydrogen column densities the effects of ionization are likely to be much more important. Nonetheless, this 
is certainly an area worth investigation. 

The ionisation levels in sub-DLAs are still not particularly well known, which is why we have presented a broad range of cases to illustrate the 
range of possible values in determining $\Omega_{met}$ and $\rho_{Z}$. With further UV spectra of sub-DLAs, the higher ionization lines of S III, Si III, 
Fe III could be accessed, providing further constraints on
the levels of ionization in the gas. The Si III/Si II ratio seems to be a promising ionization indicator, as the ratio increases faster than other 
adjacent ion ratios such as Fe III/Fe II, enabling tighter constraints to be place on the ionization parameter. 
Simultaneously, the H$_2$ transitions could also be observed to determine the molecular content in these systems.  

This series of papers has roughly doubled the total number of sub-DLA observations, and nearly 
quadrupled the sample at $z<1.5$, and done so with quality high resolution spectra. 
Complementary to studying these galaxies in absorption, the typically more difficult task of imaging these
galaxies still needs to be undertaken. 
Deep imaging in multiple bands (g$^{\prime}$, r$^{\prime}$, i$^{\prime}$, K) of sub-DLAs is necessary to determine the stellar content of these systems. Almost certainly, 
the sightlines that produce sub-DLAs are sampling a variety of morphological types of gas rich galaxies. The systems with high metallicity and higher velocity widths may however be more 
representative of massive spiral galaxies.

%%%%%%%%%%%%%%%%%%%%%%%%%%%%%%%%%%%%%%%%%%%%%%%%%%%%%%%%%%%%%%%%%%%%%%%%%%%%%%%%%%%%%%%%%%%%%%%%
%%%%%%%%%%%%%%%%%%%%%%%%%%%%%%%%%%%%%%%%%%%%%%%%%%%%%%%%%%%%%%%%%%%%%%%%%%%%%%%%%%%%%%%%%%%%%%%%

\section*{Acknowledgments}
We thank the exceptionally helpful staff of Las Campanas Observatory for their assistance during the observing runs. We would also like to thank the 
anonymous referee for their helpful suggestions which greatly improved the presentation and substance.  
J. Meiring and V.P. Kulkarni gratefully acknowledge support from the National Science 
Foundation grant AST-0607739 (PI Kulkarni). J. Meiring acknowledges partial support from a South Carolina Space Grant graduate student fellowship for
a portion of this work.

\bsp

\label{lastpage}

\end{document}